\documentclass[runningheads]{llncs}
\usepackage[T1]{fontenc}
\usepackage{graphicx}
\usepackage[firstpage]{draftwatermark}

\usepackage{listings}
\usepackage[dvipsnames]{xcolor}
\usepackage{pifont}
\usepackage{color}
\usepackage{mathtools}
\usepackage[T1]{fontenc}  
\usepackage[scaled=0.7]{beramono}  
\usepackage{bm}
\usepackage{pgf,tikz}
\usepackage{pgffor}
\usepackage{pgfplots}
\usepackage{graphicx}
\usepackage{caption}
\usepackage{subcaption}
\usepackage{adjustbox}
\usepackage{microtype}
\usepackage{booktabs}
\usepackage{multirow}
\usepackage{colortbl}
\usepackage{realboxes}
\usepackage{paralist}
\usepackage{algorithm}
\usepackage[noend]{algpseudocode}
\usepackage{amsmath,mleftright}
\usepackage{physics}
\usepackage{marvosym}
\usepackage{longtable}
\usepackage[most]{tcolorbox}
\usepackage{relsize}
\usepackage{tabularx}

\usepackage{xspace}
\newcommand{\ginger}{\textsc{Ginger}\xspace}

\usepackage[backgroundcolor=Apricot,disable]{todonotes}

\definecolor{codegray}{rgb}{0.5,0.5,0.5}
\definecolor{darkgray}{rgb}{0.33,0.41,0.47}
\definecolor{backcolour}{rgb}{0.96,0.96,1.0}
\lstdefinestyle{mystyle} {
	language=Java,
	backgroundcolor=\color{backcolour},   
	basicstyle=\small\ttfamily,
	numbers=left,                    
	numbersep=5pt,     
	numberstyle=\tiny\color{codegray},
	tabsize=2,
	breaklines=true,
	alsoletter=-,
	commentstyle=\color{darkgray},
	morecomment=[l]{;},
	literate=
	{;pre-conditions:}{{{\color{darkgray};pre-conditions:}}}{15}
	{;instant}{{{\color{darkgray};instantVoltage = maxVoltage * cos(2 * pi * frequency * time)}}}{62}
	{public}{{{\textbf{public}}}}{6}
	{ensures}{{{\textbf{ensures}}}}{7}
	{assignable}{{{\textbf{assignable}}}}{10}
	{requires}{{{\textbf{requires}}}}{8}
	{diverges}{{{\textbf{diverges}}}}{8}
	{loop\_invariant}{{{\textbf{loop\_invariant}}}}{14}
	{normal\_behavior}{{{\textbf{normal\_behavior}}}}{15}
	{also}{{{\textbf{also}}}}{5}
	{\\forall}{{{\color{MidnightBlue}\@backslashchar{}forall}}}{7}
	{\\result}{{{\color{MidnightBlue}\@backslashchar{}result}}}{7}
	{\\strictly\_nothing}{{{\color{MidnightBlue}\@backslashchar{}strictly\_nothing}}}{17}
	{\\fp_nan}{{{\color{OliveGreen}\@backslashchar{}fp\_nan}}}{7}
	{\\fp_nice}{{{\color{OliveGreen}\@backslashchar{}fp\_nice}}}{8}
	{\\fp_infinite}{{{\color{OliveGreen}\@backslashchar{}fp\_infinite}}}{12}
	{\\fp_normal}{{{\color{OliveGreen}\@backslashchar{}fp\_normal}}}{10}
	{fp.gt}{{{\color{MidnightBlue}\textbf{fp.gt}}}}{5}
	{FloatingPoint}{{{\color{MidnightBlue}\textbf{FloatingPoint}}}}{13}
	{fp.isNaN}{{{\color{MidnightBlue}\textbf{fp.isNaN}}}}{8}
	{fp.isInfinite}{{{\color{MidnightBlue}\textbf{fp.isInfinite}}}}{13}
	{fp.div}{{{\color{MidnightBlue}\textbf{fp.div}}}}{6}
	{fp.sub}{{{\color{MidnightBlue}\textbf{fp.sub}}}}{6}
	{fp.mul}{{{\color{MidnightBlue}\textbf{fp.mul}}}}{6}
	{fp.add}{{{\color{MidnightBlue}\textbf{fp.add}}}}{6}
	{RNE}{{{\color{MidnightBlue}\textbf{RNE}}}}{3},
	morekeywords = [2]{object, def, Real, val, require, set-logic, declare-fun, assert, check-sat},
	escapeinside={(*@}{@*)}
}
\usepackage[hidelinks]{hyperref}
\usepackage{orcidlink}
\let\orcidID\orcidlink

\newtcolorbox{rqanswer}[1][]{
    colback=gray!10!white,
    colframe=gray!50!black,
    boxrule=1.2pt,
    arc=2mm,
    left=1pt,
    right=1pt,
    top=0pt,
    bottom=2pt
}
\usepackage{framed}

\definecolor{sidebargrey}{gray}{0.6}

\newenvironment{greybar}{%
  \MakeFramed{\advance\hsize-\width \FrameRestore}%
}%
{\endMakeFramed}

\newif\ifarxiv
 \arxivtrue

\begin{document}
\SetWatermarkText{\hspace*{3.7in}\raisebox{8.55in}{\href{https://eapls.org/pages/artifact_badges/}{\mbox{\includegraphics[width=1.4cm]{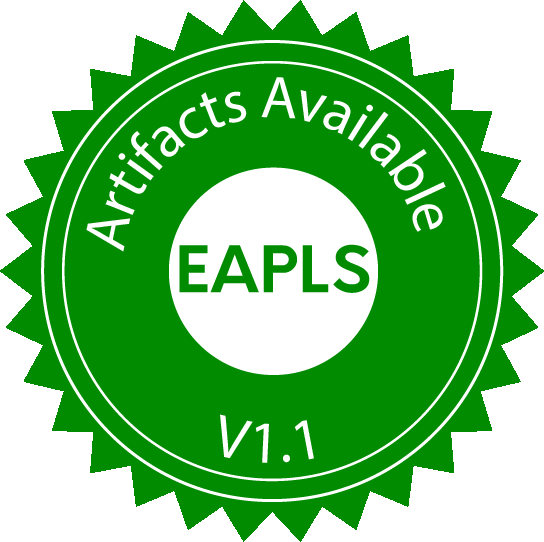}\includegraphics[width=1.4cm]{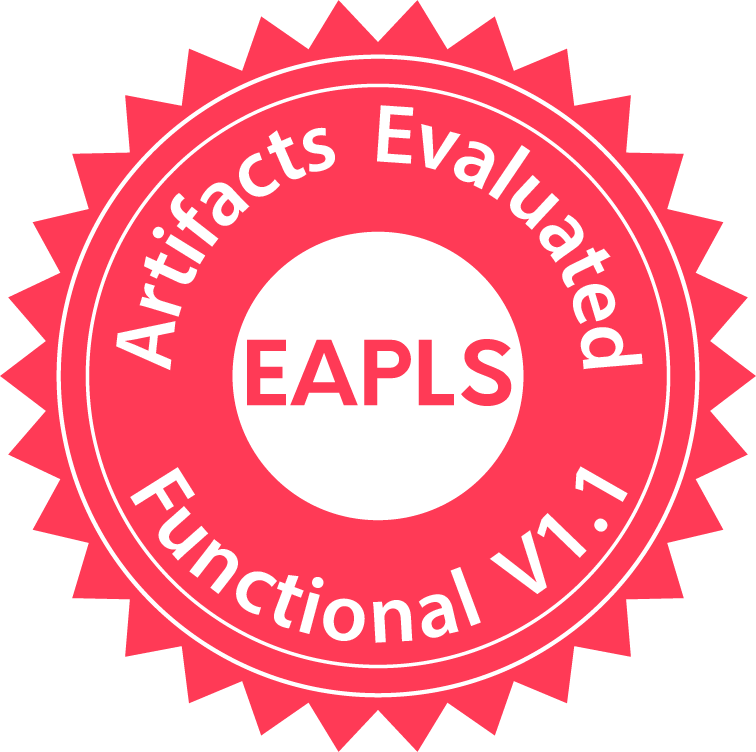}}}}}
\SetWatermarkAngle{0}
\title{Metamorphic Testing for Floating-Point Performance Issues in SMT Solvers}
%
%
\author{Rosa Abbasi\inst{1}\orcidID{0000-0003-1495-3470} (\Letter)
 \and Eva Darulova\inst{2}\orcidID{0000-0002-6848-3163} }
\authorrunning{R. Abbasi et al.}
%
\institute{MPI-SWS, Kaiserslautern and Saarbr\"{u}cken, Germany, \email{rosaabbasi@mpi-sws.org} \and
Uppsala University, Uppsala, Sweden, \email{eva.darulova@it.uu.se}
}
\maketitle              
\begin{abstract}
SMT solvers are essential in various domains, including program verification and synthesis.
Although their correctness and performance have been extensively
studied, performance testing for the floating-point theory remains limited, particularly
for real-world queries.
%
We propose a metamorphic testing approach that uses semantics-preserving rewrite
rules, focusing on floating-point special values and peep-hole optimizations, to uncover meaningful
performance issues in SMT solvers’ handling of floating-point formulas. Using
real-world test inputs, our approach is able to identify for every solver tested SMT
queries for which solving time increases when the queries are simplified;
we see slowdowns of up to 33.4x for Z3, 5.6x for MathSAT, 4.5x for cvc5, and 1.8x for Bitwuzla.
We further observe that the approach is less successful when using input files that were
randomly generated from a grammar.
 
\end{abstract}

\section{Introduction}





For the last few decades, SMT solvers have been used in many domains, such as
program verification~\cite{KeYBook2016,detlefs2005simplify}, symbolic
execution~\cite{cadar2008klee,godefroid2005dart}, and program
synthesis~\cite{alur2013Synthesis}.
The efficiency and correctness of tools that use SMT solvers highly depend on
the solver's efficiency and correctness, so the solvers have been subjected
to extensive correctness
testing~\cite{yao2021fuzzing,storm2020,winterer2020unusual,yao2021skeletal,murxla2022,sun2023last,xia2024fuzz4all} and,
on fewer occasions, performance testing~\cite{scott2020banditfuzz,ET2024}.
Given the complexity and scale of modern solver implementations,
SMT solvers are typically treated as black boxes during testing,
with issues being detected and reported to solvers' developers
rather than debugged at the implementation level.
As a result, SMT solvers have significantly improved over time across various
theories.

However, performance on queries over the floating-point theory remains expensive
and unpredictable; we observed unexpected performance behavior in
our prior work~\cite{AbbasiEtAl2021}. This is unfortunate, as automated verification
support for  floating-point arithmetic specifically is highly valuable
as it is unintuitive to many developers.
For example, reasoning about
floating-point special values (Not-a-Number (\texttt{NaN}), infinity, or signed
zeroes) is challenging and often misunderstood~\cite{dinda2018developers}.


To illustrate an instance of unexpected performance behavior, consider the
partial SMT query shown in \autoref{listing-running-example-before} that
we extracted from one of the verification conditions (VCs) generated by the
KeY deductive verifier~\cite{KeYBook2016} for a real-world Java program. The property that KeY checks
in this case is that scaling a rectangle does not result in \texttt{NaN} nor
infinity.
To keep the listing brief, we exclude the remaining assertions covering other
verification-condition branches, as well as the variable definitions and
assertions regarding their range constraints.
Z3, cvc5,
MathSAT, and Bitwuzla require 9.37, 1.98, 3.4, and 0.25 seconds (average across 5 runs), respectively
to solve this SMT query and produce \texttt{unsat}.
		\begin{lstlisting}[caption = Partial SMT Query generated by KeY, label=listing-running-example-before, basicstyle=\footnotesize\ttfamily]
	; additional details elided, incl. variable declaration and range constraints
	(assert (fp.leq  (fp.sub RNE  (fp.add RNE  (fp.add RNE
	(fp.mul RNE  |field_benchmarks.rectangle.Rectangle::$x| ui_arg1)
	(fp.mul RNE  |field_benchmarks.rectangle.Rectangle::$y|
	(fp #b0 #b0...0 #b00...00))) (fp #b0 #b0...0 #b00...00)) ; constants 0.0 (zeroes elided)
	(fp.add RNE  (fp.add RNE  (fp.mul RNE
	(fp.add RNE  |field_benchmarks.rectangle.Rectangle::$x|
	|field_benchmarks.rectangle.Rectangle::$width|) ui_arg1)
	(fp.mul RNE  |field_benchmarks.rectangle.Rectangle::$y|
	(fp #b0 #b0...0 #b00...00))) (fp #b0 #b0...0 #b00...00)))	(fp #b0 #b0...0 #b00...00)))
		\end{lstlisting}



We now \emph{simplify} the multiplication terms on lines 4 and 9.
We use the floating-point identity twice that if the first operand in a multiplication is negative and neither
\texttt{NaN} nor infinity and the second operand is either \texttt{$+0.0$} or \texttt{$-0.0$},
then the multiplication can be replaced by the negation of the second operand:
\begin{equation*}
	u_1 \neq (\texttt{NaN} \lor \pm\infty)\ \land\ u_1<0~\land\ (u_2 = +0.0 \lor u_2 = -0.0)
	\Rightarrow
	(u_1 \times u_2) \longrightarrow -u_2
\end{equation*}


Having simplified the query, we would expect improved solver performance. This
is, however, only the case for Z3 and cvc5, whose solving times improve by 22\%
and 7\% (7.30s, 1.84s), respectively. For MathSAT and Bitwuzla, the solving time
increases by 28\% and 40\% (to 4.37s and 0.35s).
We are specifically interested in  slowdowns in solving time and call them performance issues.



Our goal is to identify such \textbf{performance issues} in the implementation of the floating-point
theory
across various SMT solvers on real-world and grammar-generated benchmarks.
We define a performance issue in this context as a statistically significant slowdown \emph{of an individual solver} on a semantically equivalent, structurally simpler query, while acknowledging that such transformations may interact differently with solver heuristics and preprocessing.


Performance issues in floating-point reasoning affect multiple stakeholders.
Identifying solver-specific performance issues can help SMT solver developers to
debug and optimize their implementations.
Tool developers who rely on SMT solvers can benefit
from insights into how floating-point constraints can be effectively simplified
before they are passed to SMT solvers, e.g., by optimizing VCs.
End users of such tools may be
discouraged from enabling or using floating-point verification support when it results in
excessive and unpredictable runtimes, despite the much-needed robust floating-point support in
verification tools~\cite{hahnle201724}.

Detecting such performance issues in SMT solvers is non-trivial, and has not been considered
in existing testing approaches that primarily focus on correctness~\cite{yao2021fuzzing,storm2020,winterer2020unusual,yao2021skeletal,murxla2022,sun2023last,xia2024fuzz4all} rather than
performance~\cite{scott2020banditfuzz,ET2024}. Those that consider performance record timeouts,
or compare solving times between different solvers, i.e., they rely on oracles or perform differential testing.
Neither of those is well-suited to detect performance issues for an individual solver. Defining an oracle would
amount to defining the expected solving time, which is not feasible.
Comparing different solvers against each other evaluates a different metric;
a longer runtime compared to different solvers \emph{on an individual SMT file} does not necessarily
indicate a performance issue in our sense, rather these individual differences may stem
from different internal heuristics, implementation choices, or algorithms.
While comparing different solvers (on the same inputs) is certainly important,
we believe that identifying unexpected behavior of one solver on (highly)
related inputs provides orthogonal and important insights.



We propose metamorphic testing to detect performance issues. Starting from an
input SMT query using the floating-point theory, our approach applies up to two
semantics-preserving rewrite rules (randomly chosen from a larger set), each of
which aims to \emph{simplify} the query---we would thus expect the query to be
solved faster. By comparing individual solver performance on the original vs.
rewritten queries, we can identify issues when the rewritten query
takes longer to solve.
Rule application is limited to a maximum of two rules per file to balance introducing meaningful syntactic variation with avoiding excessive simplification, which can mask performance issues by consistently producing speedups.\footnote{Once the identified performance issues are addressed, each rewrite rule on its own is expected to consistently improve solver performance.}

As a secondary objective, we can compare performance sensitivity to these
rewrite rules across solvers to further highlight the unpredictability of the
floating-point theory implementations in SMT solvers.
%
In contrast to prior metamorphic testing targeting incompleteness bugs~\cite{Bringolf2023}, we use semantics-preserving rewrites to detect unexpected performance slowdowns within individual SMT solvers.
While we focus on the floating-point theory due to its unpredictable
performance and unintuitive nature, the approach itself generalizes to other SMT theories.


Our rewrite rules are drawn from common simplifications in floating-point
reasoning and include rules targeting floating-point special values (Not-a-Number (NaN),
infinities, signed zeroes), as well as rules derived from
floating-point compiler (peep-hole) optimizations~\cite{notzli2016lifejacket}.
While many of these rules involve special values, which are known to be
challenging for developers~\cite{dinda2018developers}, all rules are semantics-preserving and aim to
simplify floating-point expressions.
Our implementation in the prototype tool \ginger applies such rewrite rules to
SMT queries and compares solver performance before and after rewriting fully
automatically.

An additional challenge, particularly for floating-point arithmetic, is that
floating-point SMT queries stemming from real-world benchmarks are rare. Only
few floating-point benchmarks in SMT-COMP have any special values,
and when they do, those test files are usually quite small.
As we show in this paper, automatically generating floating-point input
files, e.g., from a grammar, with semantic and structural complexity of
real-world problems is non-trivial, too.

We apply our approach to four SMT solvers (cvc5, Z3, MathSAT, Bitwuzla) using
three benchmark suites: verification conditions from real-world programs, SMT queries (randomly) generated from a grammar, and mutations derived
from real-world verification conditions.
Our results show that our metamorphic testing approach is effective in detecting
performance issues. We are able to identify performance issues in every solver
tested: we observe slowdowns on the simplified files of up to 33.4x for Z3, 5.6x
for MathSAT, 4.5x for cvc5, and 1.8x for Bitwuzla. We also observe that
realistic test inputs, along with mutants derived from them, are significantly
more effective in exposing performance issues compared to grammar-generated test
files.

As expected (and illustrated in the example above),
the rewrite rules affect different solvers differently, i.e., simplifying
(real-world) SMT queries improves solving time for some, but typically not for
all tested solvers.
Moreover, files that undergo more rewrite-rule applications tend to show larger
improvements, highlighting the potential of rewriting---whether as a preprocessing
step or within solver implementations---as a practical performance optimization.

\paragraph{Contributions} In summary, our main contributions are:
\begin{itemize}
	\item We present a metamorphic testing–based approach for detecting performance issues in SMT solvers on the floating-point theory (\autoref{sec:metamorphic}).
	\item We implement this approach in the tool \textsc{Ginger} (\autoref{sec:implementation}), which is released as open source\footnote{\url{https://github.com/rosaAbbasi/ginger}, artifact: \url{https://doi.org/10.5281/zenodo.21500107}}), together with all tested SMT files.
	\item We use \textsc{Ginger} to test four SMT solvers with support for the floating-point theory on three benchmark suites (\autoref{sec:benchmark-selection}), and identify performance issues (\autoref{sec:evaluation}).
	We have submitted selected performance issues to the corresponding SMT solver developers. For cvc5, the reported performance issue was confirmed~\cite{CVC5Issue}; an update independent of this paper fixes this particular issue.
\end{itemize}

\section{Metamorphic Testing for Performance Issues}\label{sec:metamorphic}

Metamorphic testing addresses both the input generation and
the oracle problem in software testing~\cite{10.1145/3143561}.
Given a known relationship between multiple
inputs to a system, metamorphic testing defines the expected relationship between their
corresponding outputs. A violation of this expected relation indicates a potential issue.

In our context, the relationship between the original and rewritten SMT formulas is one of
semantic equivalence and simplification, achieved through the application of a set of semantics-preserving
floating-point rewrite rules to the input SMT files. Since both versions
represent the same logical problem and the rewrite rules are designed to
\emph{simplify} formulas, we expect solvers to produce identical
results (\texttt{sat}/\texttt{unsat}) and the rewritten formula to have
comparable or reduced solving time.
Here, we do not use a formal notion of simplification; rather, simplification
refers to transformations that reduce the number of nodes in SMT expressions
and, from the perspective of users inspecting SMT queries, result in
structurally simpler expressions.

Metamorphic testing thus fits naturally for identifying performance issues in SMT
solvers, as it enables evaluating each solver in isolation without the need for
an external performance oracle or reference solver---both of which are difficult
to define in this context. 

\begin{figure}[t]
	\centering
	\includegraphics[width=0.9\textwidth,alt={Flowchart showing how the input SMT file is parsed into an AST, normalized, and rewritten twice. If any rewrite rule was applied, both the input and rewritten SMT file are solved and the solving time and result is recorded in a CSV file.}]{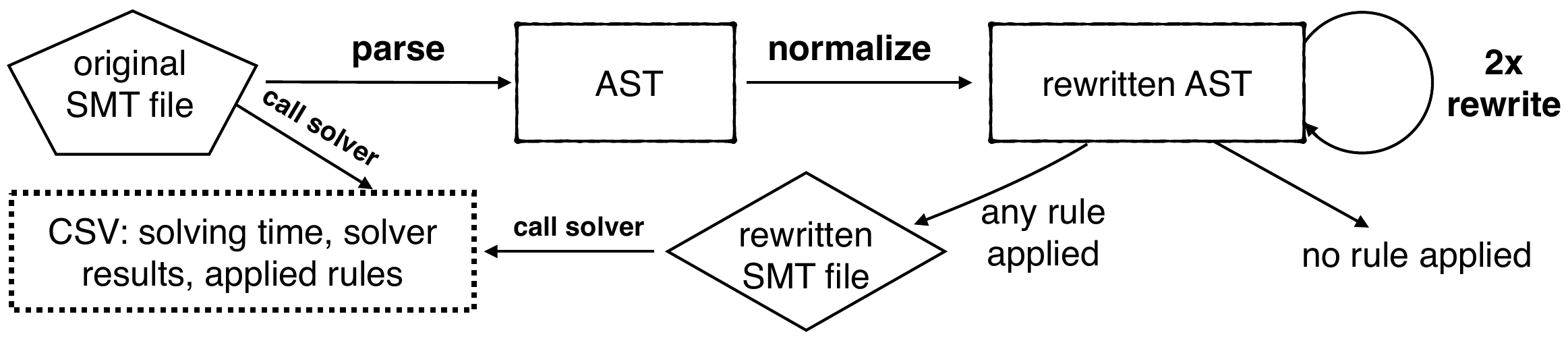}
	\caption{Workflow of our metamorphic testing approach for performance issues}
	\label{fig:workflow}
\end{figure}
\autoref{fig:workflow} illustrates the workflow of our metamorphic testing approach.
Each SMT test input file is parsed into an abstract syntax tree (AST) and
normalized slightly to ensure easier applicability of the rules.
A randomly shuffled set of rewrite rules is applied sequentially to all
\texttt{assert} statements (in all of our benchmarks, all relevant operations
appear in \texttt{assert} statements) such that at most two rules are applied to
a file.
We limit the application to at most two rules per file to avoid oversimplifying
formulas, which would obscure performance anomalies by causing only speedups.
We observed empirically that using (at most) two rules works consistently well across our benchmarks.
We additionally shuffle the rule order so that, under this limit, we vary which rules
get applied.

Each
\texttt{assert} statement is processed recursively, with applicable rules applied to any
nested expression.
Meaning, a rule may be applied multiple times within a file, depending on the number of
matching terms.
If at least one rule is applied to a test file, all selected solvers
are executed five 
times on both the original and rewritten version of the file.
We record all statistically significant relative differences in runtime between these two versions.
Regressions, i.e.{} when the rewritten file is solved slower, indicate a potential performance issue.

Note that the order of rule application matters, as we impose a limit of at most
two rule applications per file. In particular, the first two rules (possibly the same rule applied twice) whose preconditions are satisfied are applied,
even though additional rules might also be applicable to the same file.
Consequently, different rule application orders generate different transformed
files. Moreover, applying one rule modifies the query, and depending on which
rule is applied, the preconditions for another rule may no longer hold, thereby
preventing its application.
While the concrete transformations may differ depending on the order of rule application, we  experimented with different rule orders and found that the overall trends observed in our large-scale experiments remain stable, with no qualitative differences in the results.



\subsection{Normalization}
Before applying the rewrite rules, we perform a syntactic normalization step to
ensure that special-value handling is explicit and uniform across all
benchmarks. These transformations
improve the consistency of
subsequent rewriting and analysis steps and simplify the implementation.
These transformations are themselves a form of rewrite rule application;
thus, the term normalization is used informally here and should not be understood in the
formal sense in term rewriting.
This preprocessing  pass  is applied once and includes the following transformations:

\renewcommand{\theenumi}{\arabic{enumi}}
\begin{enumerate}
	\item  Occurrences of  values such as \texttt{$+0.0$}, \texttt{$-0.0$}, \texttt{NaN}, \texttt{$-\infty$}, and \texttt{$+\infty$} that appear in assertions of the form \texttt{(assert (fp.eq [variable] [value]))} or \texttt{(assert (= [variable] [value]))} are replaced by substituting the variable with the corresponding constant throughout all assertions.
	\item Additional assertions are introduced to explicitly state that a constant is not \texttt{NaN}, not infinite, and not zero, whenever its assigned floating-point value does not correspond to these special cases.
	\item  When a variable is constrained by upper and lower bounds—expressed through comparisons such as \texttt{fp.gt} or \texttt{fp.lt} with constant floating-point values— if applicable, new assertions are added to declare that the variable is not zero, not infinite, not \texttt{NaN}, and that it has the appropriate sign (positive or negative).
	\item  Equality checks such as \texttt{(=u (\_ NaN [eb] [sb]))} or \texttt{(=(\_ NaN [eb] [sb]) u)} are replaced with the semantically equivalent predicate \texttt{(fp.isNaN u)}.
\end{enumerate}

\subsection{Rewrite Rules}

The rewrite rules are semantics-preserving transformations over the SMT-LIBv2
floating-point theory.
Although expressed as rewrite rules, they are not used as a term-rewriting
system and are not intended to reach a normal form. 
In total, our rule set consists of 51 rewrite rules that we derived from three sources:
\renewcommand{\theenumi}{\Roman{enumi}}
\begin{enumerate}
	\item the IEEE 754-2008 standard for floating-point arithmetic~\cite{ieee75408},
	\item verified compiler optimizations from LifeJacket~\cite{notzli2016lifejacket}, and
	\item additional rules we inferred for handling special values.
\end{enumerate}

We collected the rules in group I directly from identities from
the IEEE standard.
For group II, we adapted rule implementations from the LifeJacket repository
to the SMT context, ignoring rules that are not relevant to SMT formulas such as optimizations relying on the \texttt{nsz} flag, which ignores the sign of
zero.
Finally, we inferred several rules from recurring special-value checks observed in floating-point verification conditions. These rules simplify predicates such as \texttt{fp.isNaN}, \texttt{fp.isInfinite}, and \texttt{fp.isZero} when applied to compound expressions by exploiting semantic information about their operands. 
The rules are designed primarily for simplification, normalization, and redundancy reduction in \texttt{assert} statements.
We verified all rules with SMT solvers before including them in the rule set. Alternatively, rewrite-rule frameworks such as Rare and IsaRare could be used
for the specification and formal verification of SMT rewrite rules~\cite{Noetzli2022,lachnitt2024isarare}.

\begin{table}[t]
	\centering
	\caption{ A subset of rewrite rules}
	\label{tab:rules}
	\fontsize{8}{10}\selectfont  
	\begin{tabular}{p{3.2cm}p{7.5cm}c}
		\hline
		\textbf{Name} & \textbf{Description} & \textbf{Source} \\
		\hline
		ieee-NaN-2-3-5-6 & ($\Rightarrow u+\texttt{NaN} \longrightarrow \texttt{NaN}), (\Rightarrow \texttt{NaN} + u \longrightarrow \texttt{NaN}), (\Rightarrow u\cross\texttt{NaN} \longrightarrow \texttt{NaN}), (\Rightarrow \texttt{NaN}\cross u \longrightarrow \texttt{NaN})$ & I \\
		ieee-infinity-19-20 & $(\Rightarrow -(+\infty)\longrightarrow -\infty),( \Rightarrow -(-\infty)\longrightarrow +\infty)$ &  I\\
		ieee-NaN-7 &$\Rightarrow -\texttt{NaN} \longrightarrow \texttt{NaN}$&  I\\
		ieee-infinity-36-39-43-46 & $(u=0 \Rightarrow u\cross+\infty\longrightarrow\texttt{NaN}), (u=0 \Rightarrow +\infty\cross u\longrightarrow\texttt{NaN}), (u=0 \Rightarrow u\cross-\infty\longrightarrow\texttt{NaN}), (u=0\Rightarrow -\infty\cross u\longrightarrow\texttt{NaN}) $&  I\\
		simplify-7 & $(\Rightarrow u - (+0.0) \longrightarrow u),( \Rightarrow u -( -0.0) \longrightarrow u)$ &  II\\
		simplify-1 & $(\Rightarrow u + (-0.0) \longrightarrow u), (\Rightarrow  u + (+0.0) \longrightarrow u) $& II \\
		simplify-1-2 & $(\Rightarrow -0.0 + u \longrightarrow u), (\Rightarrow +0.0 + u \longrightarrow u)$ & II  \\
		addSub-6 & $u_1 - (-0.0 - u_2) \longrightarrow u_1 + u_2$ & II \\
		isNaN-add-1 & $u_2 \neq (\texttt{NaN} \land \pm\infty)\Rightarrow ( u_1 + u_2 ) = \texttt{NaN}  \longrightarrow u_1 = \texttt{NaN}$ & III \\
		isNaN-add-2 & $u_1 \neq (\texttt{NaN} \land \pm\infty) \Rightarrow ( u_1 + u_2 ) = \texttt{NaN}  \longrightarrow u_2 = \texttt{NaN} $& III \\
		isNaN-mul-1 &$ u_2 \neq (\texttt{NaN} \land \pm\infty\land \pm0.0)  \Rightarrow ( u_1 \cross u_2 ) = \texttt{NaN} \longrightarrow u_1 = \texttt{NaN} $&  III\\
		isNaN-mul-2 & $u_1 \neq (\texttt{NaN} \land \pm\infty\land \pm0.0) \Rightarrow ( u_1 \cross u_2 ) = \texttt{NaN} \longrightarrow u_2 = \texttt{NaN}$&  III\\
		isZero-mul-1 & $u_2 \neq (\texttt{NaN} \land \pm\infty)  \land u_2<0 \land  (u_1 = +0.0 \lor u_1 = -0.0) \Rightarrow u_1 \cross u_2 \longrightarrow -u_1$ &  III\\
		isZero-mul-2 & $u_1 \neq (\texttt{NaN} \land \pm\infty)  \land u_1 <0 \land  (u_2 = +0.0 \lor u_2 = -0.0) \Rightarrow u_1 \cross u_2 \longrightarrow -u_2 $&  III\\
		\hline
	\end{tabular}
\end{table}
\autoref{tab:rules} lists a subset of rewrite rules---those most frequently applied on our
benchmarks---along with their descriptions and sources of derivation. The complete list of rules is provided in
\ifarxiv
Appendix \autoref{sec:appendix-1}.
\else
Section A.1 in the appendix of the extended version of this paper~\cite{TechReport}.
\fi
Some entries correspond to multiple
related transformations grouped
under a single rule; in
such cases, each transformation is enclosed in parentheses
and separated by commas.
Each transformation is written as a logical implication: the expression to the left of
$\Rightarrow$ specifies the preconditions that must hold for the transformation to
apply, while the equation on the right defines the actual rewrite, where the left-hand
side of the "$\longrightarrow$" is replaced by the right-hand side.
The variables $u$, $u_i$, denote floating-point variables of type  
\texttt{Float32} or \texttt{Float64}, though one can generalize the rules to other
precisions.
In the rule syntax, the symbol “$\land$” denotes conjunction, “$\lor$” denotes
disjunction, and “\texttt{RNE}” in the precondition indicates that the rounding mode
must be round-to-nearest-even.
When a precondition uses $\pm\infty$ or $\pm0.0$ (e.g., $u \neq \pm\infty$), it
abbreviates both signs: $u \neq \pm\infty$ means $(u \neq +\infty)\land(u \neq
-\infty)$, and analogously $u \neq \pm0.0$ means $(u \neq +0.0)\land(u \neq -0.0)$.
For example, the rule \texttt{isNaN-add-1} states that if the floating-point variable
$u_2$ is neither NaN nor infinity then the expression
$(\texttt{fp.isNaN}~(\texttt{fp.add}~\texttt{RNE}~u_1~u_2))$ can  be replaced with
$(\texttt{fp.isNaN}~u_1)$. 

\section{Benchmark Selection and Generation}\label{sec:benchmark-selection}

Our empirical results (somewhat unsurprisingly) show that selecting relevant test
inputs is important. Since most of our rewrite rules include special values in their preconditions or transformations, we need
input files that exercise the floating-point theory including special values.

SMT-COMP is a common source of SMT benchmarks, however, this collection
contains relatively few instances involving special values or explicit queries
about them: out of 43,076 files from the non-incremental subset, only 5,067
mention any IEEE-754 special values,
and only 62 of these allowed any rewrite rules to be applied. Furthermore, the solving times are so
small that for cvc5 and Bitwuzla not a single file has a non-trivial solving time.
Therefore, we exclude SMT-COMP benchmarks from our main
evaluation.

Instead, we focus on an existing real-world benchmark set, and attempt to generate additional
input formulas from scratch from a grammar, as well as by mutating the existing files\footnote{All benchmark suites are publicly available at \url{https://github.com/rosaAbbasi/ginger/tree/main/benchmarks}.}.

\subsection{KeY-Derived Verification Conditions (K\textsmaller{E}Y-FP)}
Our first suite consists of verification conditions related to the floating-point theory
generated by the KeY deductive verifier from realistic code~\cite{AbbasiEtAl2021}.
We chose these instances because some of the checked properties focus on
the absence or controlled use of special values, among other functional properties.
We use the benchmarks as they were released.

\subsection{Grammar-Generated Floating-Point Benchmarks  (GG\textsmaller{EN}-FP)}
We generated the second suite by extending the BanditFuzz SMT fuzzing framework~\cite{scott2020banditfuzz} and creating grammar-based floating-point seed files with this
extension. Concretely, we added some missing floating-point constructs, i.e., the special
values \texttt{NaN}, positive and negative infinity,
positive and negative zero, and the predicate \texttt{fp.isZero}. We also re-implemented a
weight parameter that biases the generator toward particular constructs as a substitute
for the default uniform random selection.

To obtain a higher fraction of formulas that involve or test for special values, we adjust
the generation probabilities of floating-point constructs across three batches while
keeping the structural parameter ranges fixed. In all batches, we reduce the probability
of generating fused multiply–add (FMA) operations to 0.01 and eliminate the use of
round-nearest-away (RNA) rounding mode, since MathSAT does not support them. We also
double the probability of predicates related to special-value checks, namely
\texttt{fp.isNaN}, \texttt{fp.isInfinite}, and \texttt{fp.isZero}. In the second batch, we
build upon the first configuration by further doubling the generation probability of basic
arithmetic operations (\texttt{fp.add}, \texttt{fp.mul}, \texttt{fp.sub}, and
\texttt{fp.div}). In the third batch, we extend this setup by additionally doubling the
probability of generating special floating-point values (\texttt{NaN} etc.)

We vary three structural parameters for the generated formulas: (i) the number
of assertions, (ii) the depth of assertions, and (iii) the number of variables.
For each parameter, we select the minimum value of 5 and maximum value of 15 and
instantiate the Cartesian product over these ranges, and repeat the generation
process five independent times. Since BanditFuzz introduces randomness, repeated runs
result in different benchmarks.
We randomly select the precision for each case to be either \texttt{Float32} or \texttt{Float64} uniformly.


During our empirical study, we observed that among the 16,877 generated files
with at least one applied rewrite rule, many were trivial. Across 67,548
corresponding solver runs, the overall average runtime was only 0.05 seconds,
and just 72 instances exhibited an average runtime exceeding 10 seconds. That is, 
most generated formulas were too simple to meaningfully
challenge the solvers. We thus only use the 72 instances
with longer runtimes for actual solver testing.

\subsection{Mutations of  K\textsmaller{E}Y-FP Instances  (K\textsmaller{E}YM\textsmaller{UT}-FP)}
Our third benchmark suite was motivated by the difficulty of generating
realistic instances using the grammar-based approach. This suite consists of
mutations  of 12 selected test files from
the \textsc{KeY-FP} benchmark suite. Before applying mutations, we manually
removed uninterpreted sorts and all assertions related to heap and Java class
structures, retaining only the pure floating-point components. This
simplification ensures compatibility with Bitwuzla since it does not support
the theory of integers and equalities
over uninterpreted sorts.

To generate mutants, we used the seed-based mutation testing feature of
Sparrow~\cite{yao2021skeletal}, setting the theory to \texttt{QF\_FP} and configuring it
to generate 100 mutants per seed file. Since the original implementation of
Sparrow did not support saving all generated mutants, we modified the code to
enable this functionality. The produced mutants were neither semantically nor
syntactically preserving. We filtered out all mutants that did not conform to the
SMT-LIBv2 grammar and caused solvers to produce syntax errors.

\section{Implementation}\label{sec:implementation}
Our prototype, \ginger, is implemented in Python and communicates with off-the-shelf SMT
solvers by
generating SMT-LIBv2 files and invoking their command-line interfaces.
\ginger implements the entire workflow described in \autoref{sec:metamorphic}.
\ginger applies the rewrite rules, illustrated in \autoref{tab:rules},
to the SMT-LIB
abstract syntax tree. The system is designed to be modular—new rules can be
incorporated simply by extending the rule list. \ginger
applies the rules
sequentially in the order they appear and recursively traverses each
\texttt{assert} statement to ensure that all applicable transformations are
performed. 
For each benchmark instance, \ginger logs the set of applied rules for
subsequent analysis.

\paragraph{Recorded Metrics}
\ginger records solver runtimes on the original benchmarks and then re-measures these runtimes
after applying the normalization and rewriting. Runtimes are measured using the bash \texttt{time} command.
To account for runtime variability, each solver–benchmark combination is executed five times,
with a timeout of 300 seconds.
Consequently, for each file and solver, we obtain two sets of five runtime measurements:
one set before the rewriting and one set after.
For each benchmark file and solver, we  compute the relative difference for each of the
five runs as
$r_i= (t_{\texttt{before}} - t_{\texttt{after}})/t_{\texttt{before}}$
where $t_{\texttt{before}}$ and $ t_{\texttt{after}}$ denote the runtimes before and
after rewriting for run $i$. Positive values of $r_i$ indicate a speedup, whereas negative values indicate a
slowdown. This yields five relative difference values per solver per file, capturing run-to-run variability in
performance change.

In addition to solver runtimes, \ginger records the \emph{normalization time} and \emph{rewrite
	time} (in seconds) for each instance, 
measuring the overhead introduced by preprocessing and transformation. 
For each solver, we also log the result on both the original and rewritten files (i.e.,
\texttt{sat}, \texttt{unsat}, \texttt{unknown}, \texttt{timeout}, or \texttt{error}), 
as well as the set  of rewrite rules applied to each instance.

\section{Empirical Results}\label{sec:evaluation}
We test the SMT solvers cvc5, Z3, MathSAT, and Bitwuzla with our approach.
Our focus is on the following research questions:

\begin{itemize}
	\item[\textbf{RQ1:}] \emph{How does rewriting affect solver performance on real-world, grammar-generated, and mutation-based benchmarks?}
	This RQ captures the per-solver effects of rewriting with at most two rules across different benchmark suites.
	\item[\textbf{RQ2:}]\emph{How does rewriting affect the solver performance across the different
	solvers?}
	This RQ investigates cross-solver variability.
	\item[\textbf{RQ3:}] \emph{How does applying the full set of semantic-preserving rewrite rules on real-world and mutation-based test files affect performance?}
	This RQ investigates the potential of using (unrestricted) rewriting for performance optimization.
\end{itemize}



\subsection{Experimental Setup}\label{sec:expreimental-setup}

We run our experiments on a server with 256GB
memory and 2x8 CPU cores at 3.3GHz. However, \ginger
runs in a single thread and does not use more than 8GB
of memory. We use four SMT solvers: Z3~\cite{10.1007/978-3-540-78800-3_24} (version 4.15.2), cvc5~\cite{barbosa2022cvc5}
(version 1.3.0), MathSAT~\cite{mathsat5} (version 5.6.11) and Bitwuzla~\cite{10.1007/978-3-031-37703-7_1} (built from
latest GitHub commit~\cite{bitwuzla2025commit}). We run all solvers using a single thread.

As described in the implementation section, we compute five relative difference
values per solver per file, capturing run-to-run variability in performance
change. These values serve as the input data for our statistical analysis, where
we perform a two-sample t-test at a significance level of  $\alpha = 0.05$ to
determine whether the mean relative difference differs significantly from zero, under the null hypothesis that it is equal to zero (i.e. no statistically significant slowdown or speedup).

When test cases are too small or trivial, i.e.{} yield a runtime
within $\epsilon=10^{-2}$ of
zero, they are excluded from analysis. Consequently, only
SMT files
with a non-zero baseline (before-rewriting) runtime are considered
\emph{testable} cases.
Among these, we categorize benchmarks that exhibit statistically significant
relative
differences across all solvers as \emph{uniformly significant benchmarks}. We
classify
those that show significance for only a subset of solvers as \emph{partially
	significant}.
The remaining benchmarks, for which the null hypothesis cannot be rejected by
any solver,
are referred to as \emph{uniformly insignificant}.
These three categories are mutually exclusive. 

We consider the three benchmark sets \textsc{KeY-FP},
\textsc{KeYMut-FP} and \textsc{GGen-FP} described in~\autoref{sec:benchmark-selection}.
We do not run Bitwuzla on the \textsc{KeY-FP} set, since Bitwuzla does not support the theory of integers and equalities over uninterpreted sorts that these files contain.
\begin{table}[t]
	\centering
	\caption{Median and average runtimes per solver across benchmark suites}\label{tab:runtimes}
	\fontsize{8}{10}\selectfont  
	\renewcommand{\arraystretch}{1.0}
	\setlength{\tabcolsep}{10pt}
	\begin{tabular}{@{}lllllll@{}}
		\toprule
		\multirow{2}{*}{Solver} & \multicolumn{2}{c}{\textsc{KeY-FP}} & \multicolumn{2}{c}{\textsc{KeYMut-FP}} & \multicolumn{2}{c}{\textsc{GGen-FP}} \\
		& avg.   & median  & avg.  & median  & avg.  & median  \\
		\midrule
		Z3      & 111.91 & 97.21 & 18.82 & 10.19 & 0.18  & 0.01  \\
		cvc5    & 1.22   & 1.06  & 1.65  & 1.39  & 1.74  & 1.33  \\
		MathSAT & 1.76   & 1.40  & 1.83  & 1.79  & 27.83 & 19.38 \\
		Bitwuzla & \texttt{n/a} & \texttt{n/a} & 0.16 & 0.17 & 0.35 & 0.27 \\
		\bottomrule
	\end{tabular}
\end{table}
For each solver, \autoref{tab:runtimes} reports the median and average runtime
on original files (i.e., before rewriting), considering only those files to
which at least one rule was applied. As shown, for each benchmark suite, there is at
least one solver with noticeably longer runtimes, indicating that the test files
are non-trivial and that solvers' runtimes vary substantially, supporting
our decision to compare each solver against itself rather than against
other solvers.

\subsection{RQ1: Rewriting Effect on Different Benchmark Sets}




\begin{table}[t]
	\caption{Statistically significant changes in solving time after rewriting}
	\label{tab:stats}
	\resizebox{\textwidth}{!}{%
		\renewcommand{\arraystretch}{1.1}
		\centering
		\begin{tabular}{lccccccccccc}
			\toprule
			\multirow{3}{*}{\begin{tabular}[c]{@{}c@{}}Benchmark \\Suite\end{tabular}} &
			\multirow{3}{*}{\begin{tabular}[c]{@{}c@{}}\#Test \\ Files\end{tabular}} &
			\multirow{3}{*}{\begin{tabular}[c]{@{}c@{}}\#Uniformly\\ Sig.\end{tabular}} &
			\multirow{3}{*}{\begin{tabular}[c]{@{}c@{}}\#Uniformly\\ Insig.\end{tabular}} &
			\multicolumn{8}{c}{\#Partially Sig.}                                                                                                                                    \\ \cline{5-12}
			&                                                                          &                                                                             &                                                                               & \multicolumn{2}{c}{Z3}                      & \multicolumn{2}{c}{cvc5}      & \multicolumn{2}{c}{MathSAT}                 & \multicolumn{2}{c}{Bitwuzla}                \\
			&                                                                          &                                                                             &                                                                               & Sig.                 & Insig.               & Sig.                 & Insig. & Sig.                 & Insig.               & Sig.                 & Insig.               \\ \midrule
			\textsc{KeY-FP}                          & 64         & 40        & 2     & 17                   & 5                    & 7                   & 15     & 16       & 6                   & \texttt{N/A}        & \texttt{N/A}        \\
			\textsc{GGen-FP}                         & 72                                                                       & 3                                                                          & 31                                                                             & 0                    & 33                   & 27                   & 6      & 26                   & 7                   & 18                   & 15                    \\
			\textsc{KeYMut-FP}                         & 650                                                                       & 408                                                                          & 0                                                                             & 242                    & 0                   & 215                   & 27     & 216                   & 26                   & 34                   & 208                   \\
			\bottomrule
	\end{tabular}}
\end{table}

To identify potential performance issues, we evaluate whether rewriting leads to statistically significant
changes in solver performance, in either direction.

\autoref{tab:stats} summarizes the number of files for which rewriting shows a
statistically significant difference in solving times.
The “\#Test Files” column shows the number of files to which at least one
rewrite rule was applied; only these files are benchmarked.
\autoref{tab:stats} shows that for the \textsc{KeY-FP} and  \textsc{KeYMut-FP}
sets, rewriting leads to a statistically significant change in solving time for at least one solver
for all but a single \textsc{KeY-FP} file. Many of these changes are not uniform across solvers,
i.e.{} different solvers see a significant change for a different number of files.
For \textsc{GGen-FP}, the performance changes caused by rewriting are mostly
statistically insignificant, pointing to the simplicity of the grammar-generated test
files. 

\begin{figure}[t] 
\centering
  \begin{minipage}{\textwidth}

  \begin{subfigure}[b]{0.5\textwidth}
    \includegraphics[width=\linewidth,alt={Plot showing relative difference in solving time for individual files for MathSAT.}]{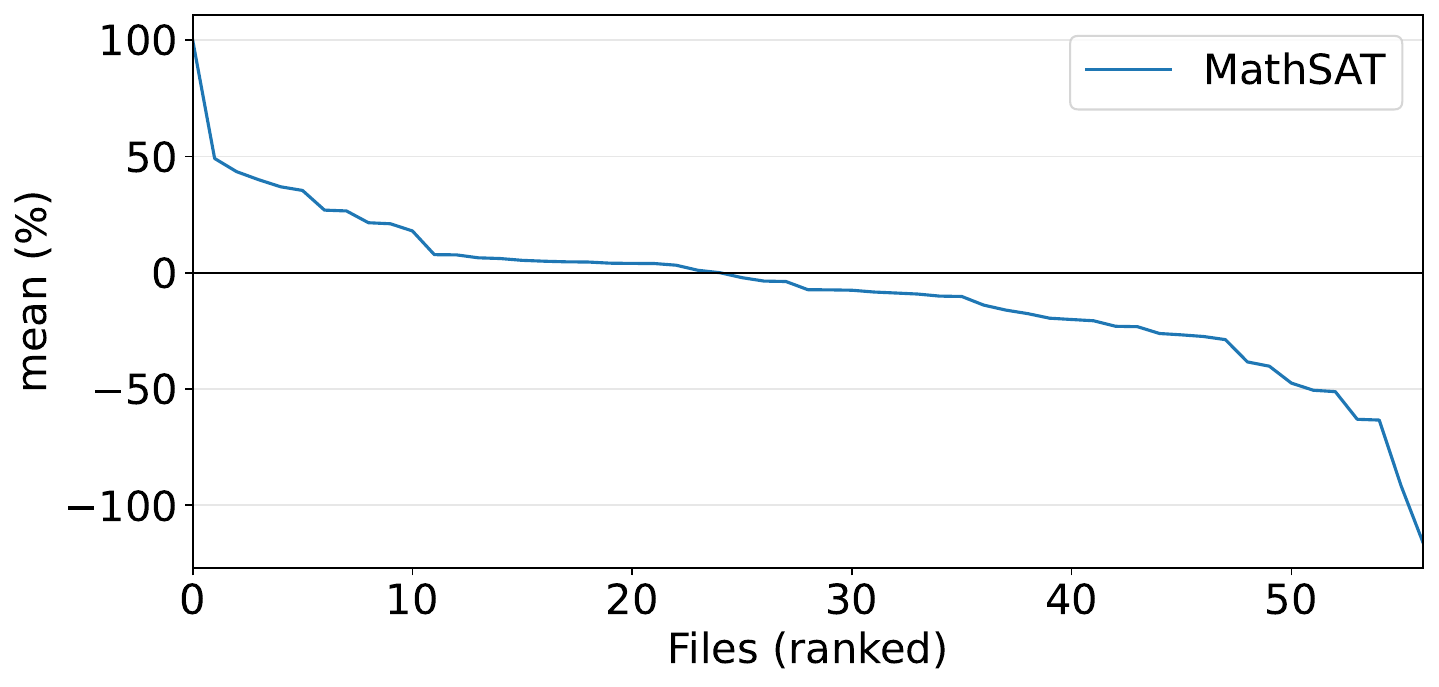}
  \end{subfigure}\hfill
  \begin{subfigure}[b]{0.5\textwidth}
    \includegraphics[width=\linewidth,alt={Plot showing relative difference in solving time for individual files for Z3.}]{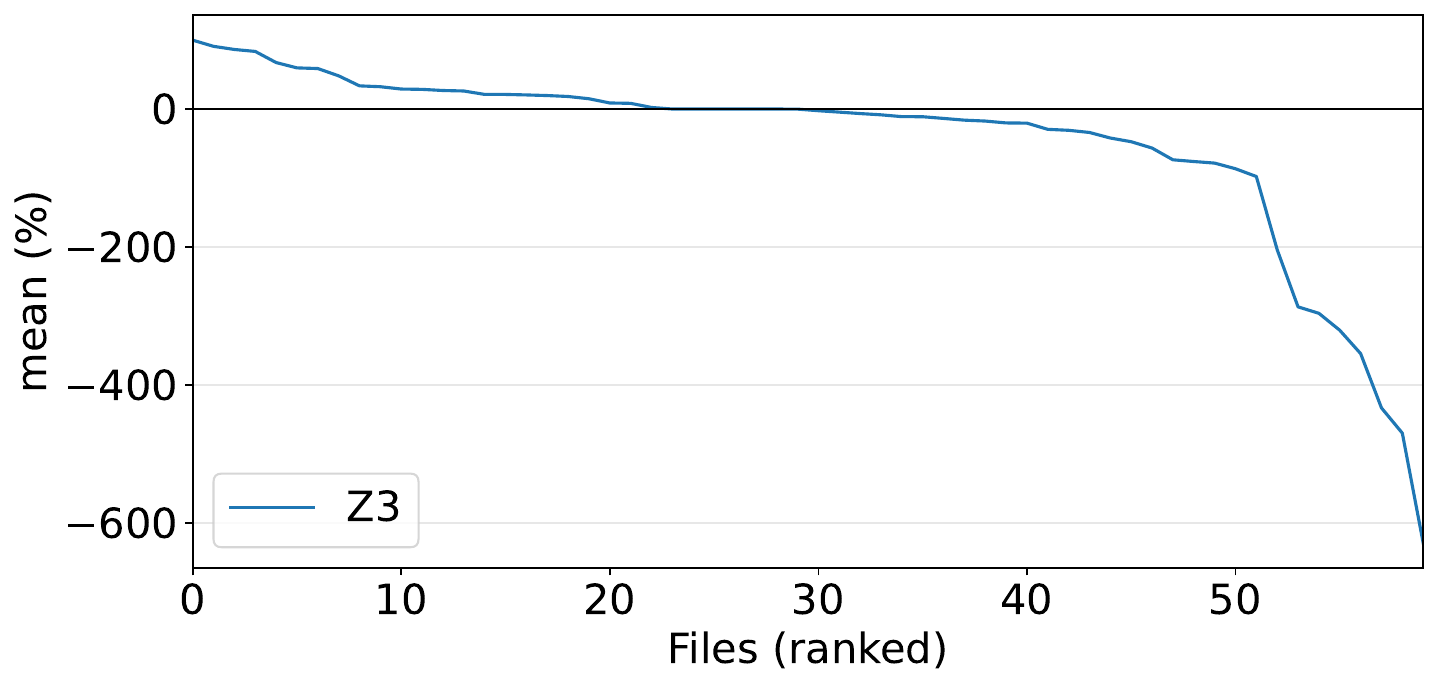}
  \end{subfigure}
  \begin{subfigure}[b]{0.5\textwidth}
    \includegraphics[width=\linewidth,alt={Plot showing relative difference in solving time for individual files for cvc5.}]{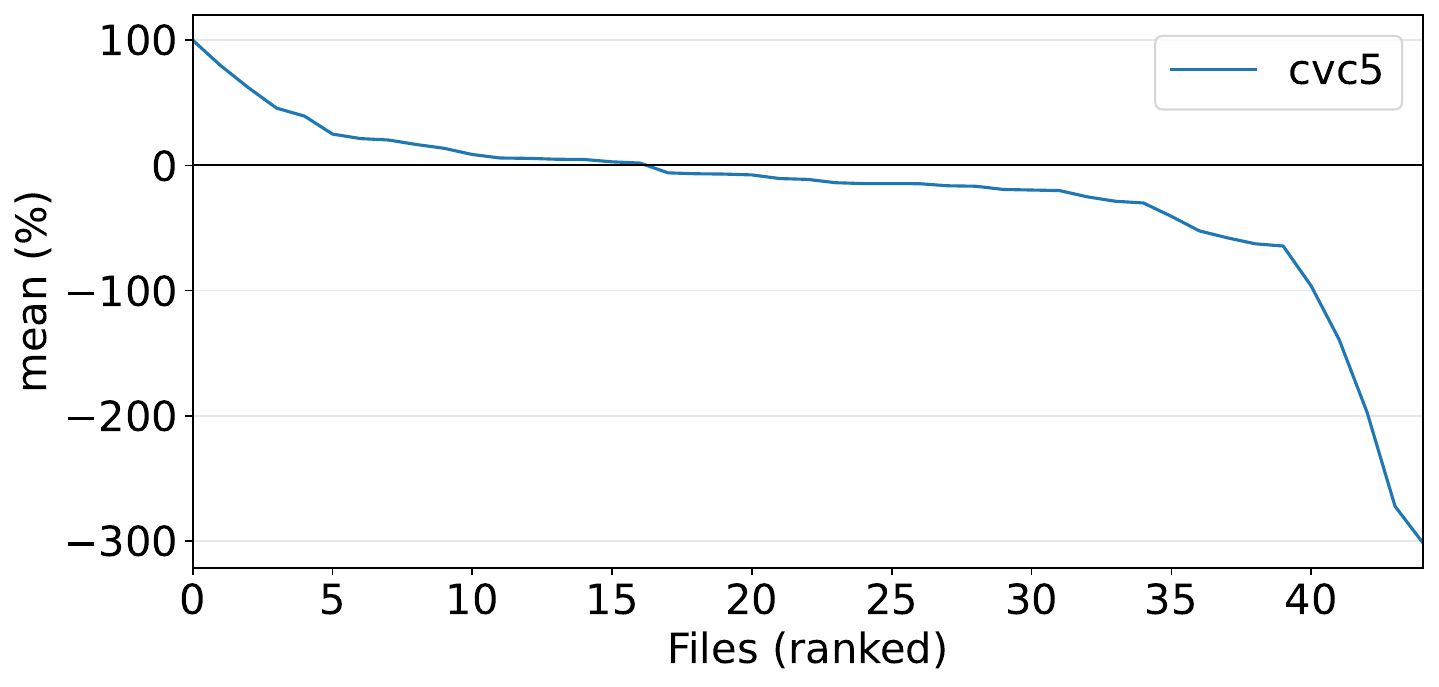}
  \end{subfigure} \hfill
  \caption{Relative difference in solving time after rewriting, \textsc{KeY-FP} suite}
  \label{fig:keyfloat-rel-improvment}
\end{minipage}
\begin{minipage}{\textwidth}
	\begin{subfigure}[b]{0.5\textwidth}
		\includegraphics[width=\linewidth,alt={Plot showing relative difference in solving time for individual files for MathSAT.}]{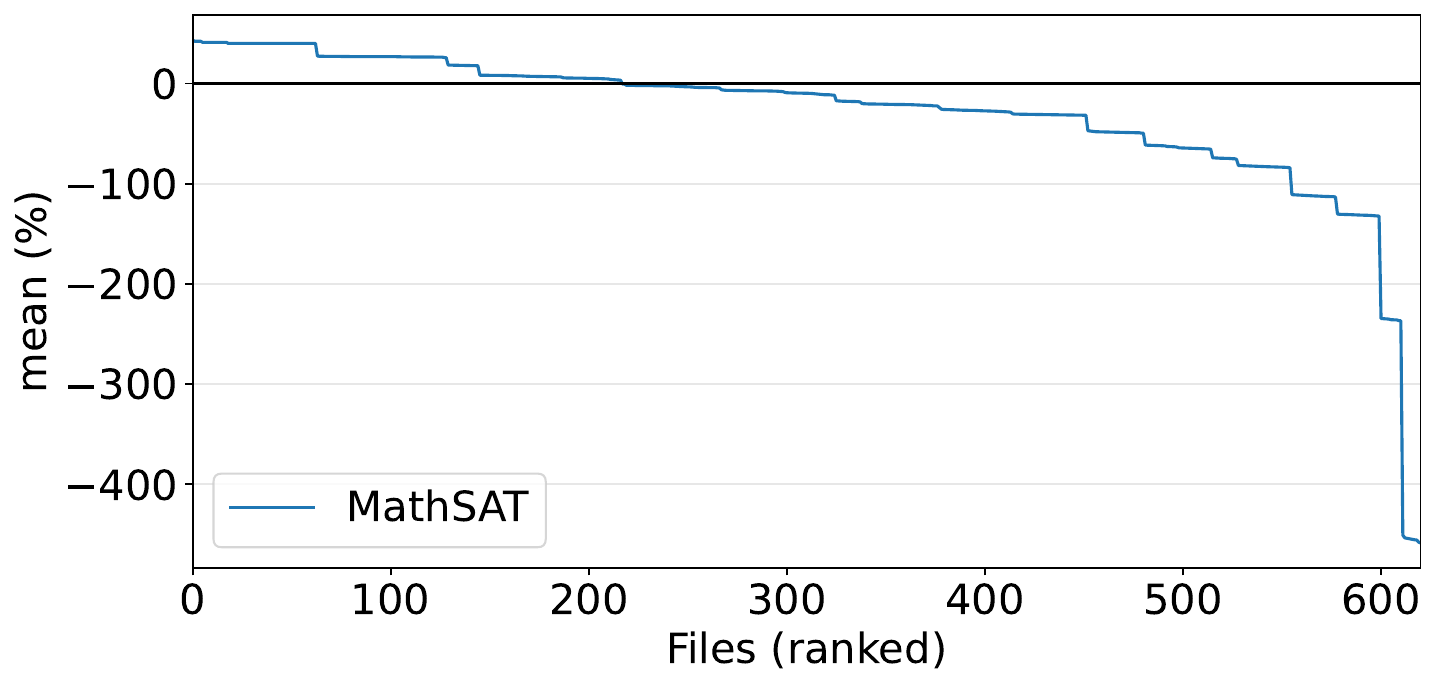}
	\end{subfigure}\hfill
	\begin{subfigure}[b]{0.5\textwidth}
		\includegraphics[width=\linewidth,alt={Plot showing relative difference in solving time for individual files for Z3.}]{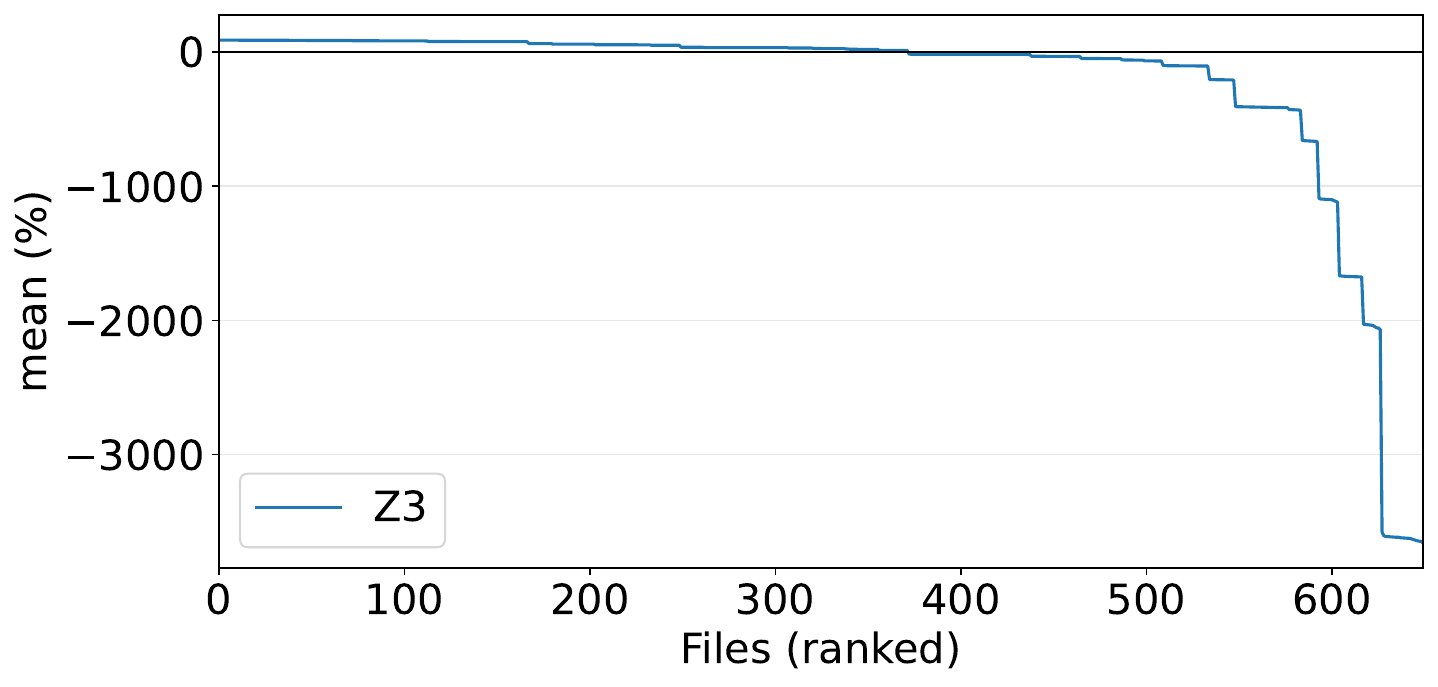}
	\end{subfigure}\hfill
	\begin{subfigure}[b]{0.5\textwidth}
		\includegraphics[width=\linewidth,alt={Plot showing relative difference in solving time for individual files for cvc5.}]{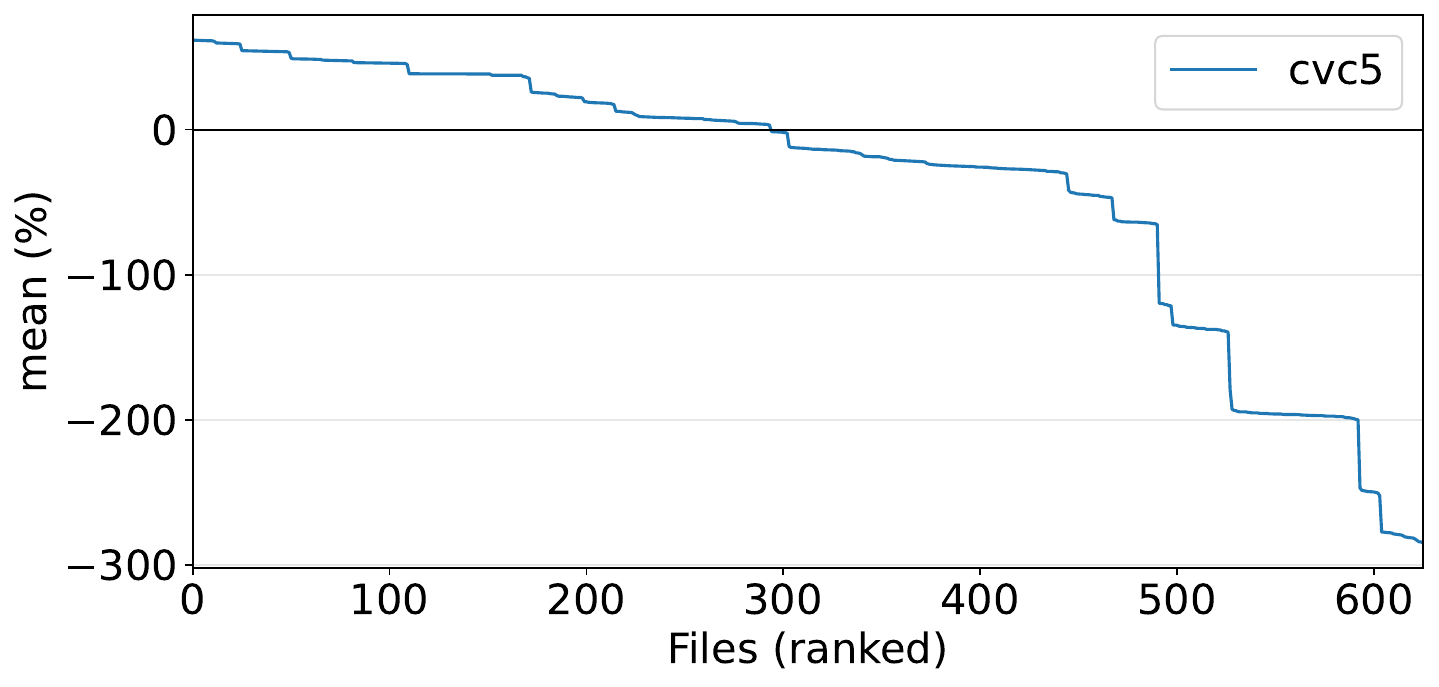}
	\end{subfigure}\hfill
	\begin{subfigure}[b]{0.5\textwidth}
		\includegraphics[width=\linewidth,alt={Plot showing relative difference in solving time for individual files for Bitwuzla.}]{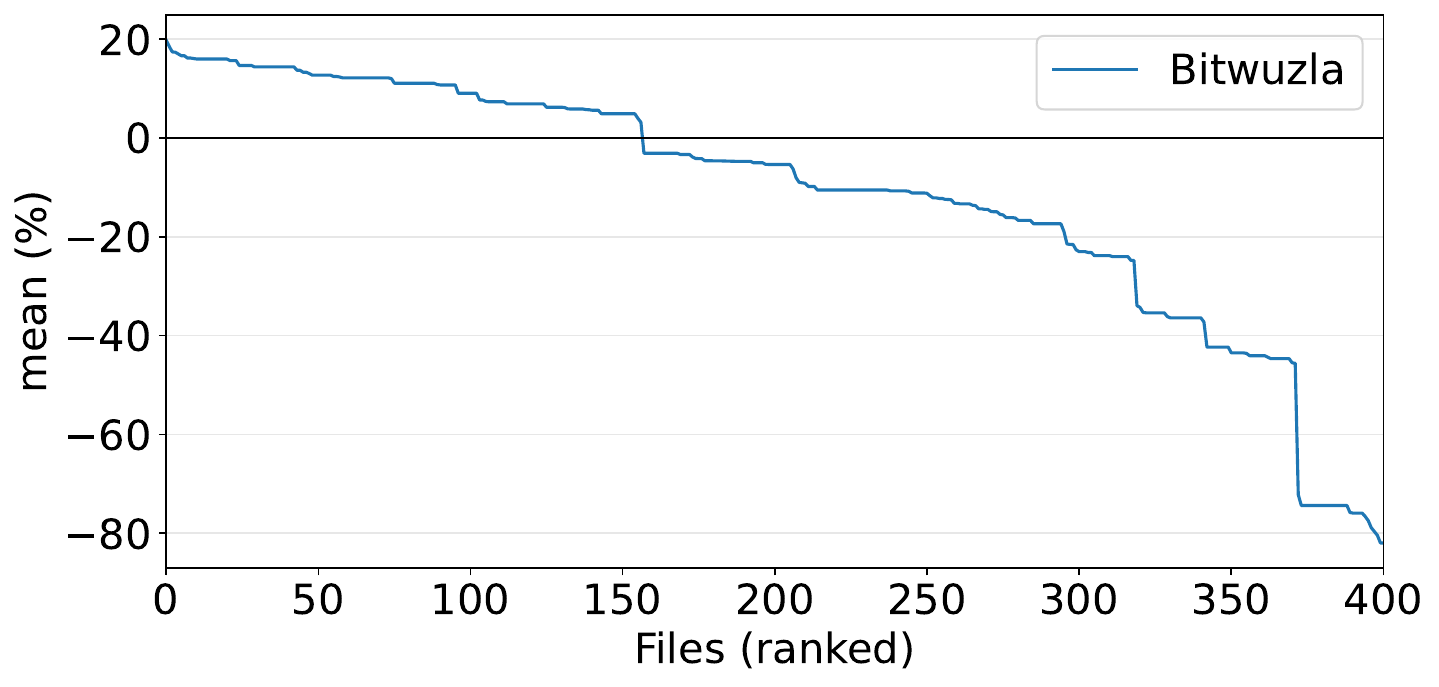}
	\end{subfigure}
	\caption{Relative difference in solving time after rewriting, \textsc{KeYMut-FP} suite}
	\label{fig:keyfloatmutants-rel-improvment}
\end{minipage}
%
\end{figure}

Figures~\ref{fig:keyfloat-rel-improvment} and ~\ref{fig:keyfloatmutants-rel-improvment}
show the magnitude of the relative
differences in solving time, for each solver sorted from the largest speedup (positive mean) to
the greatest slowdown (negative mean), for the \textsc{KeY-FP} and  \textsc{KeYMut-FP}
sets. The plots only include testable files that had a
significant difference for each individual solver (so the number of files per solver differs), and show the average over the
five runs for each file.

The observed slowdowns for the \textsc{KeY-FP} and \textsc{KeYMut-FP} benchmark
suites indicate potential performance issues.
Despite the rewrite rules being designed as simplifications, in about half of the
files they cause slowdowns, some of them of significant magnitude.

We essentially do not detect performance issues with the \textsc{GGen-FP} suite (see
\ifarxiv
Appendix Section~\ref{sec:appendix-2}
\else
Section A.2 in~\cite{TechReport}
\fi
for the magnitude of the relative differences).
For the few files that have a
statistically significant relative change, rewriting causes speedups for all
solvers---except for one file for Z3. This suggests that, in contrast to the
real-world test files in \textsc{KeY-FP} and \textsc{KeYMut-FP},
grammar-generated test files are too simple for
detecting performance issues and possibly also not representative of real-world SMT queries.

Moreover, we repeated our experiments (on \textsc{KeY-FP} and \textsc{GGen-FP}) using
different random seeds for shuffling the rule list and reanalyzed the results.
Qualitatively, the outcomes remained unchanged, indicating that our findings are
robust with respect to the applied rules.


To isolate the effect of the rewrite rules, we additionally repeated the main
experiment on the \textsc{KeY-FP} benchmark suite, but removed the extra
assertions introduced during normalization from the rewritten files. We observed
the same overall trend of speedups and
slowdowns, showing that rewrites applied to existing assertions alone are
sufficient to expose performance anomalies. We also evaluated the effect of
normalization alone by running \textsc{Ginger} on all \textsc{KeY-FP} test files
that originally had at least one rule applied. Normalization by itself produces
both speedups and slowdowns in solver runtimes. The magnitudes of these changes
are generally small, and fewer files exhibit significant differences. Since
normalization is a semantics-preserving transformation that makes implicit
information explicit or directly rewrites assertions, the slowdowns it causes
also qualify as performance issues. 
The magnitudes of the relative differences for the rewrite-only and normalization-only experiments are reported in
\ifarxiv
Appendix Sections~\ref{sec:appendix-3} and~\ref{sec:appendix-4}, respectively.
\else
Sections A.3 and A.4 in~\cite{TechReport}, respectively.
\fi


%

In the \textsc{KeY-FP} benchmarks, we also identified
7 cases for Z3 and one for MathSAT where solvers initially reported
\texttt{unsat} but timed out after rewriting. In one instance, Z3
initially timed out but reported \texttt{sat} after rewriting. We saw no mismatched
solver results in the other benchmark suites.

\paragraph{Analysis of a Reported Performance Issue.}
We submitted one representative performance issue to the developers of Z3, cvc5, and MathSAT. 
The Z3 issue received an automated response indicating a potential ``performance bug''.
For cvc5, the reported performance issue~\cite{CVC5Issue} was also acknowledged.
We further investiagted the slowdown ourselves. The slowdown persisted even
after minimizing the queries and removing auxiliary normalization assertions.
Analysis of cvc5's preprocessing and word-blasting phases revealed that the
rewrite breaks sharing of a common subexpression across assertions, causing
structurally different copies to be word-blasted separately. This cause was
also confirmed by the cvc5 developers that reproduced the reported slowdown
several weeks after paper submission. They also confirmed that the slowdown is
no longer reproducible following a SymFPU update released after the submission
of this paper.
%
This analysis highlights the value of metamorphic testing: comparing semantically equivalent queries within a single solver can reveal and help explain performance anomalies that would be difficult to identify through cross-solver comparisons.

\begin{greybar}
	\noindent{}
	\textbf{Answer to RQ1:} 
	Rewriting is effective in triggering performance slowdowns and detecting performance issues, especially when applied to real-world test files and mutations derived from them.
\end{greybar}

\subsection{RQ2: Cross-Solver Variability}
We now evaluate whether or not rewriting leads to consistent speedups or
slowdowns across different solvers.
Our goal is to evaluate how variations in floating-point theory implementations affect solvers' performance.
For this, we examine test files that had a statistically significant difference across \emph{all} solvers.

\begin{figure}[t] 
	\centering
	\begin{subfigure}[t]{0.9\textwidth}
		\begin{minipage}{0.75\linewidth}
			\centering
			\includegraphics[width=\textwidth, alt={Plot showing relatively difference of solving individual files, comparing Z3, cvc5, MathSAT on the KeY-FP benchmarks.}]{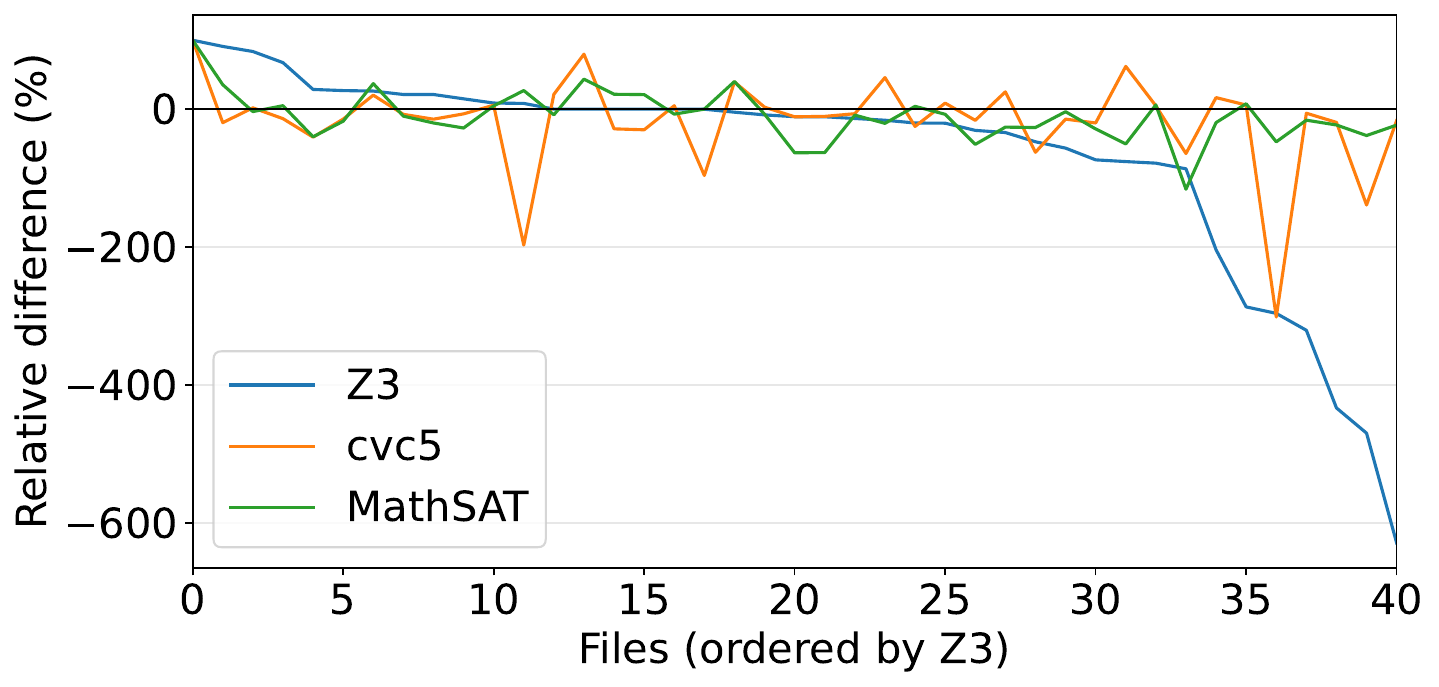}
		\end{minipage}\hfill
		\begin{minipage}{0.2\linewidth}
			\textsc{KeY-FP}
		\end{minipage}
	\end{subfigure}
	\begin{subfigure}[t]{0.9\textwidth}
		\begin{minipage}{0.75\linewidth}
			\centering
			\includegraphics[width=\textwidth, alt={Plot showing relatively difference of solving individual files, comparing Z3, cvc5, MathSAT on the KeYMut-FP benchmarks.}]{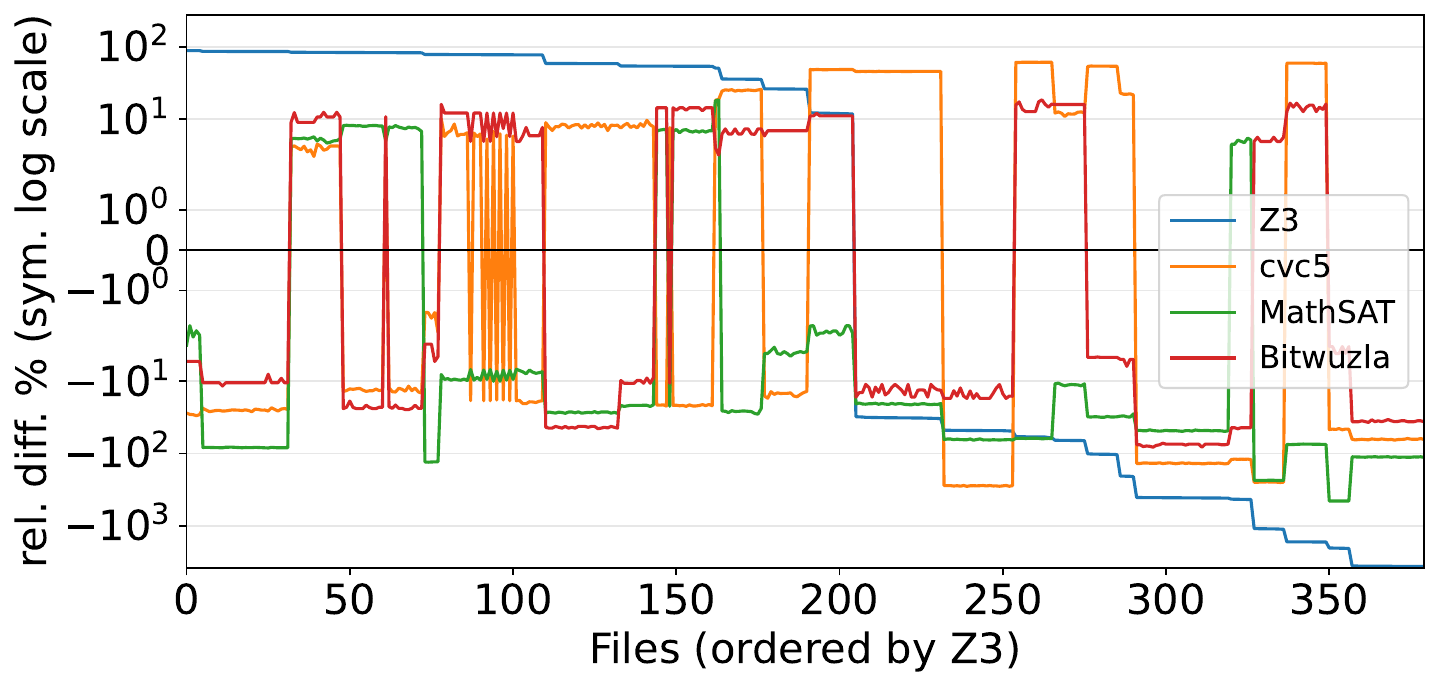}
		\end{minipage}\hfill
		\begin{minipage}{0.2\linewidth}
			\textsc{KeYMut-FP}
		\end{minipage}
	\end{subfigure}

	\caption{Comparison of relative differences in solving times across solvers}
	\label{fig:rel-impt-z3-ordered}
\end{figure}

\autoref{fig:rel-impt-z3-ordered} shows the \emph{uniformly
	significant} files from the \textsc{KeY-FP} and \textsc{KeYMut-FP} benchmark suites.
The files are
ordered according to the relative difference values for Z3. For improved
clarity, for the \textsc{KeYMut-FP} plot, we show the symmetric log-scale of
relative differences, that preserves the signs.
We excluded \textsc{GGen-FP} here because it contains only three  
uniformly significant test files, which mostly cause speedups and do not provide
interesting performance variation.

\autoref{fig:rel-impt-z3-ordered} demonstrates the unpredictability and
variability of floating-point implementations across solvers: For the same test
file and after applying the same rewrite rules, not only does the magnitude of
the relative performance change vary between solvers, but so does the direction of
the change---some solvers solve the simplified file faster and some slower. In \textsc{KeY-FP}, for 23 files,
there is a mix of slowdowns and speedups across solvers, and in the 380
uniformly significant \textsc{KeYMut-FP} files, there are 280 files
for which some solvers show a speedup while others show a slowdown.
The step-like pattern in the
\textsc{KeYMut-FP} plots arises because many mutants generated from the same
seed file differ only slightly and therefore have similar solving time before
and after rewriting.

\begin{greybar}
	\noindent{}
	\textbf{Answer to RQ2:} 
	Rewriting
	affects each solver’s performance differently, highlighting
	differences in their floating-point reasoning implementations.
\end{greybar}

\subsection{RQ3: Impact of Full Rewriting on Total Runtime}

We now evaluate whether allowing the full list of rewrite rules---without the
two-rule limit---can improve total solving time.
We compare the \textit{overall time},
i.e., the sum of runtimes for rewriting, normalization and solving the rewritten file, with the solving time of the original file.
I.e., we investigate whether solvers benefit from the simplifications produced by unrestricted rule application.
We note here that \ginger is a prototype implementation, and while designed carefully, it was \emph{not} optimized for performance.
Nevertheless, for \textsc{KeYMut-FP}, normalization and rewriting take only 0.032~s per file on average (0.034~s median).

We apply the complete set of rewrite rules—shuffled
between per-file applications and without limiting the number of applications—to
the \textsc{KeY-FP} and \textsc{KeYMut-FP} benchmark suites.
For each solver, we compute the relative differences between the original solver
runtime and the overall time (averaging solver runtimes, rewriting, and
normalization times per file over five runs).

\begin{figure}[t] 
	\centering
	\begin{minipage}{\textwidth}
		\begin{subfigure}[t]{0.47\textwidth}
			\includegraphics[width=\linewidth,alt={Plot comparing relative diffence between overall time and original solving time, comparing cvc5 and MathSAT on the KeY-FP benchmarks.}]{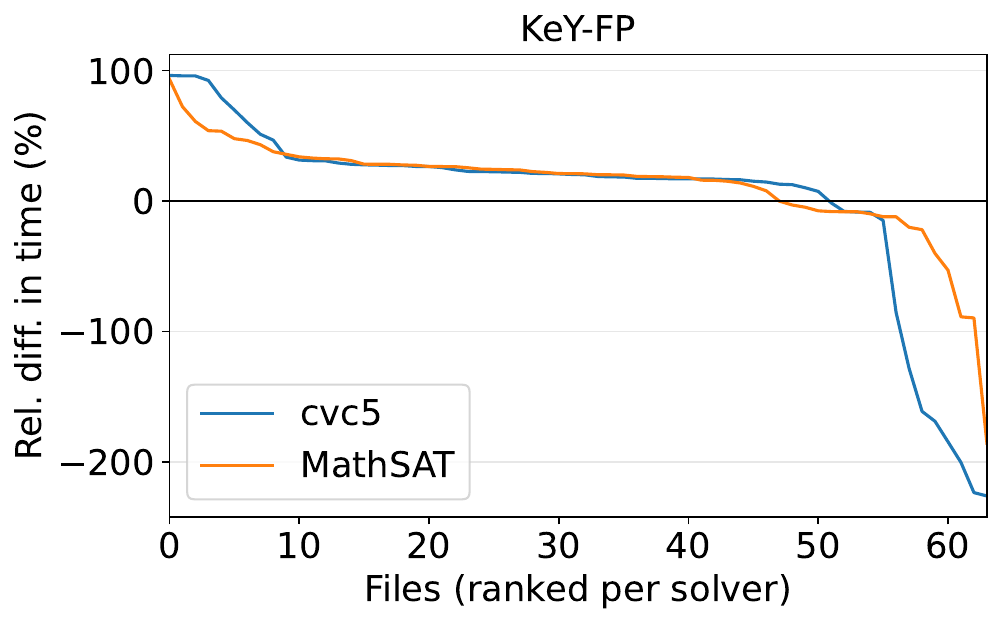}
		\end{subfigure}\hfill
		\begin{subfigure}[t]{0.47\textwidth}
			\includegraphics[width=\linewidth,alt={Plot comparing relative diffence between overall time and original solving time, showing the times for Z3 only on the KeY-fp benchmarks.}]{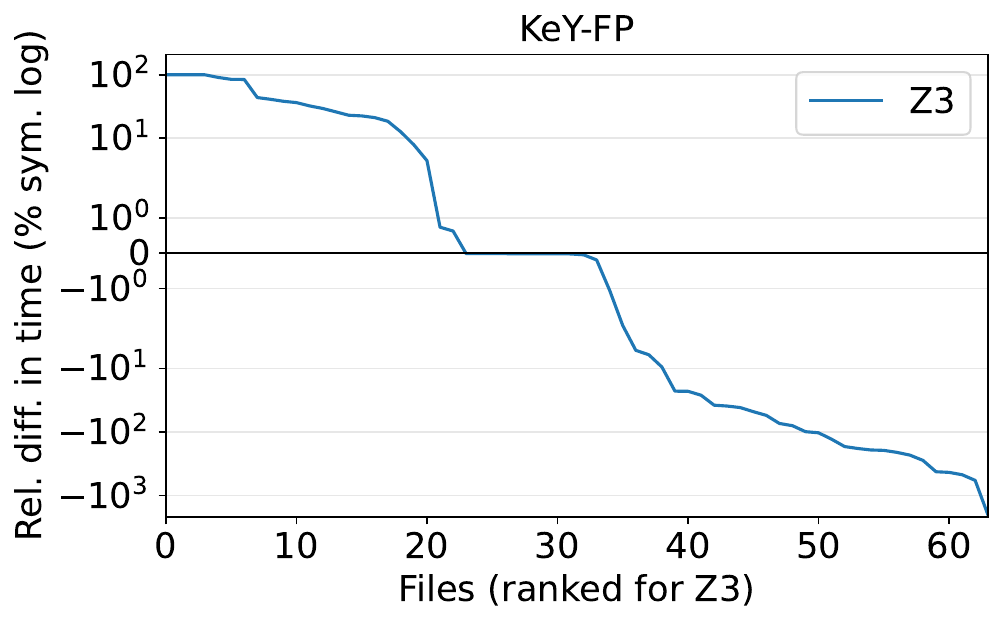}
		\end{subfigure}
	\end{minipage}
	\begin{minipage}{\textwidth}
		\begin{subfigure}[t]{0.47\textwidth}
			\includegraphics[width=\linewidth,alt={Plot comparing relative diffence between overall time and original solving time, comparing cvc5 and MathSAT on the KeYMut-FP benchmarks.}]{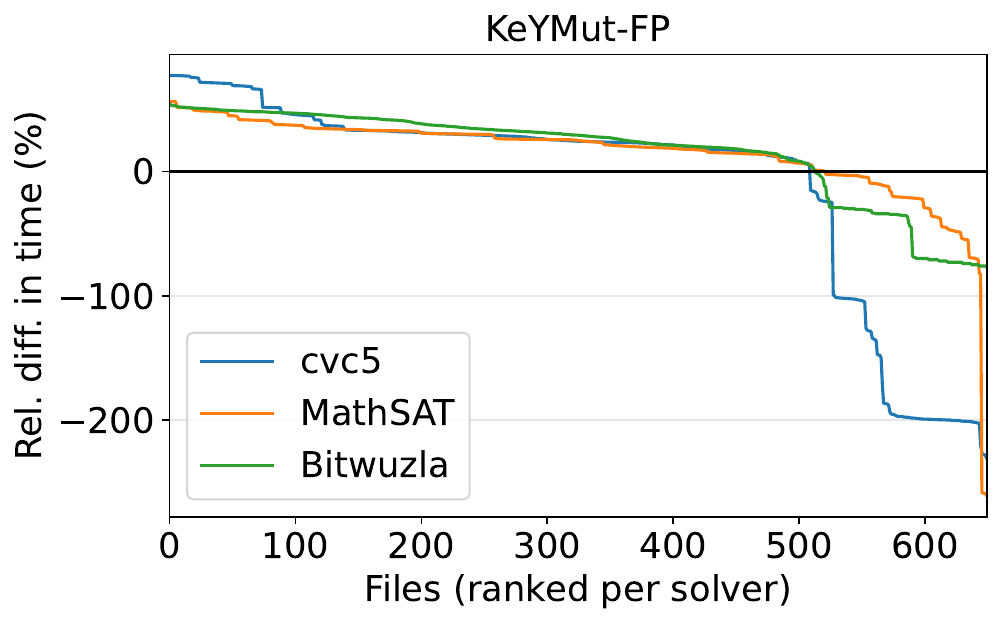}
		\end{subfigure}\hfill
		\begin{subfigure}[t]{0.47\textwidth}
			\includegraphics[width=\linewidth,alt={Plot comparing relative diffence between overall time and original solving time, for Z3 only on the KeYMut-FP benchmarks.}]{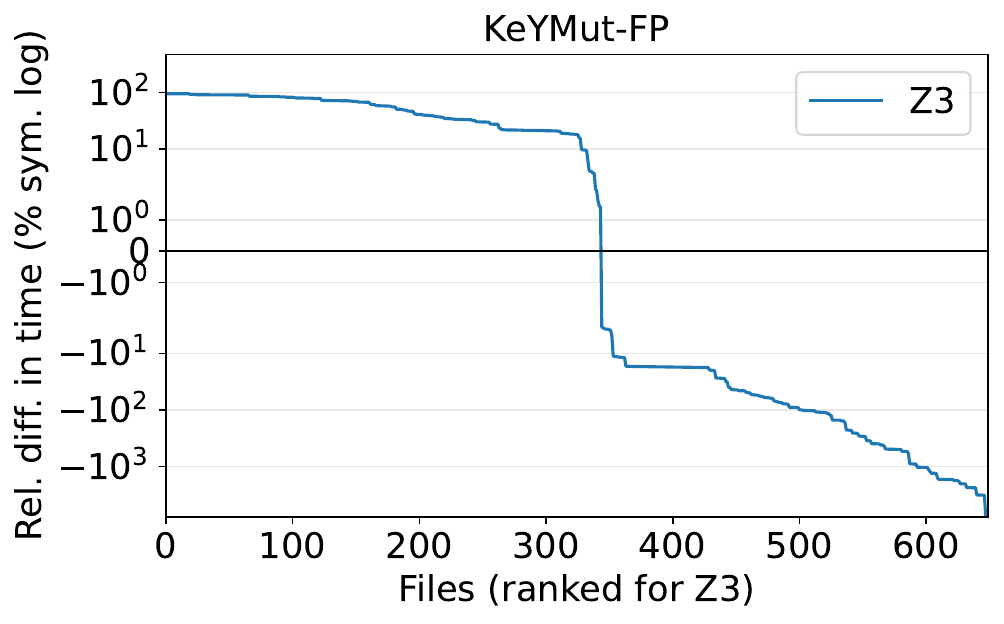}
		\end{subfigure}
	\end{minipage}
	\caption{Relative difference between overall time and original solving time}
	\label{fig:full-rule-rel-improvment}
\end{figure}
%

\autoref{fig:full-rule-rel-improvment} shows
the relative differences for the \textsc{KeY-FP} suite at the top and \textsc{KeYMut-FP} suite at the bottom. The y-axis in the Z3 plots uses a symmetric log scale.
We observe performance improvements for many input files.
For the \textsc{KeYMut-FP} files that have improved overall solving times, the median relative differences
are 48.62\% (Z3), 15.13\% (Bitwuzla), 20.74\% (cvc5) and 25.94\% (MathSAT). For \textsc{KeY-FP} files, the numbers are similar, and
we show the full summary in
\ifarxiv
Appendix \autoref{tab:pos-rel-ddif-summary}.
\else
Table 5 in~\cite{TechReport}.
\fi

\begin{greybar}
	\noindent{}
	\textbf{Answer to RQ3:} 
	Rewriting using rules such as ours may be a
	beneficial pre-processing step at least for the queries with
	floating-point special values.
\end{greybar}

\section{Related-Work}
%

SMT solvers have been subjected to extensive correctness and performance
testing as well as proof checking~\cite{stump2013smt,katz2016lazy}. Existing
testing frameworks vary in the theories they support, the classes of bugs they
target, and the techniques they employ. Several works explicitly address the
floating-point theory~\cite{storm2020,scott2020banditfuzz,murxla2022,yao2021fuzzing,winterer2020unusual,yao2021skeletal,ET2024}. Most focus on correctness, while a few
consider performance~\cite{scott2020banditfuzz,ET2024}. More recently,
Amrollahi et al.~\cite{amrollahi2025towards} identified solver stability as a
major challenge, showing that semantically equivalent SMT formulas can exhibit
substantially different runtimes due to minor syntactic variations.


BanditFuzz~\cite{scott2020banditfuzz} targets performance slowdowns using reinforcement-learning-guided fuzzing and a reference solver for performance comparison. ET~\cite{ET2024} applies grammar-based enumeration and differential testing. In contrast, our approach compares a solver against itself using semantics-preserving rewrite rules and does not rely on a reference solver or differential testing. Moreover, as we observed, grammar-generated inputs do not always yield realistic floating-point benchmarks.

%

Metamorphic testing has also been applied to floating-point SMT solving.
Storm~\cite{storm2020} generates new satisfiable formulas by recombining
subformulas from satisfiable seeds, Sparrow~\cite{yao2021skeletal} produces
equi-satisfiable variants through oracle-guided mutations, and Bringolf
et al.~\cite{Bringolf2023} use weakening and strengthening transformations.
Unlike our work, these approaches primarily target correctness rather than
performance anomalies.

Mikek et al.~\cite{Mikek2023} simplify SMT formulas using compiler
optimizations, while Pereira et al.~\cite{pereira2026smt} improve SMT
performance through simplification, normalization, and caching. Our work differs
in that it uses semantics-preserving rewrites to expose performance issues,
although we also observe that such transformations can improve solver
performance.

\section{Conclusion}

To conclude, our results demonstrate that metamorphic testing is an effective
method for detecting performance issues in SMT solvers. We have submitted
selected performance issues to the corresponding SMT solver developers,
and at least one was acknowledged, investigated, and subsequently closed.
While rewrite rules generally yield consistent
speedups on syntactically regular, generated benchmarks, their impact on
realistic benchmarks---and on mutants derived from them---is strongly solver- and
benchmark-dependent.
Our results also highlight the unpredictability of floating-point reasoning across solvers and the
influence of benchmark structure:
for some solvers, applying rewrite rules before solving
is beneficial, whereas for others it may be more appropriate to incorporate such
transformations directly into their internal heuristics.

\begin{credits}
\subsubsection{\discintname}
The authors have no competing interests to declare.
\end{credits}

\newpage
\bibliographystyle{splncs04}
\bibliography{bibliography}

\ifarxiv

\newpage
\appendix
\section{Appendix}

\subsection{All Rewrite Rules}
\label{sec:appendix-1}

\autoref{tab:all-rules} presents the complete list of rewrite rules, along with their descriptions and sources of derivation. 
Some entries correspond to multiple
related transformations grouped
under a single rule; in 
such cases, each transformation is enclosed in parentheses
and separated by commas.
Each transformation is written as a logical implication: the expression to the left of
$\Rightarrow$ specifies the preconditions that must hold for the transformation to
apply, while the equation on the right defines the actual rewrite, where the left-hand
side of the "$\longrightarrow$" is replaced by the right-hand side.
The variables $u$, $u_1$, $u_2$, and $u_3$ denote floating-point variables of type
\texttt{Float32} or \texttt{Float64}, though one can generalize the rules to other
precisions.
In the rule syntax, the symbol “$\land$” denotes conjunction, “$\lor$” denotes
disjunction, and “\texttt{RNE}” in the precondition indicates that the rounding mode
must be round-to-nearest-even.
When a precondition uses $\pm\infty$ or $\pm0.0$ (e.g., $u \neq \pm\infty$), it
abbreviates both signs: $u \neq \pm\infty$ means $(u \neq +\infty)\land(u \neq
-\infty)$, and analogously $u \neq \pm0.0$ means $(u \neq +0.0)\land(u \neq -0.0)$.
For example, the rule \texttt{isNaN-add-1} states that if the floating-point variable
$u_2$ is neither NaN nor infinity then the expression
$(\texttt{fp.isNaN}~(\texttt{fp.add}~\texttt{RNE}~u_1~u_2))$ can  be replaced with
$(\texttt{fp.isNaN}~u_1)$.

{\fontsize{8}{10}\selectfont  
	\begin{longtable}{p{2.9cm}p{8cm}c}
		\caption{List of rewrite rules}
		\label{tab:all-rules}\\
		\toprule
		\textbf{Name} & \textbf{Description} & \textbf{Source} \\
		\midrule
		\endfirsthead
		\caption[]{List of rewrite rules (continued)}\\
		\toprule
		\textbf{Name} & \textbf{Description} & \textbf{Source} \\
		\midrule
		\endhead
		addSub-1 & $(u_1 = +0.0 \lor u_1 = -0.0)\land u_2 \neq(\texttt{NaN}\land\pm 0.0) \Rightarrow u_1 + u_2 \longrightarrow  u_2 $& II \\
		addSub-2 & $(u_2 = +0.0 \lor u_2 = -0.0) \land u_1 \neq(\texttt{NaN}\land\pm 0.0) \Rightarrow u_1 + u_2 \longrightarrow  u_1$ &  II\\
		addSub-6 & $u_1 - (-0.0 - u_2) \longrightarrow u_1 + u_2$ & II \\
		ieee-NaN-2-3-5-6 & ($\Rightarrow u+\texttt{NaN} \longrightarrow \texttt{NaN}), (\Rightarrow \texttt{NaN} + u \longrightarrow \texttt{NaN}), (\Rightarrow u\cross\texttt{NaN} \longrightarrow \texttt{NaN}), (\Rightarrow \texttt{NaN}\cross u \longrightarrow \texttt{NaN})$ & I \\
		ieee-NaN-7 &$\Rightarrow -\texttt{NaN} \longrightarrow \texttt{NaN}$&  I\\
		ieee-NaN-8 & $\Rightarrow \texttt{NaN}^{-1} \longrightarrow \texttt{NaN}$ & I \\
		ieee-infinity-19-20 & $(\Rightarrow -(+\infty)\longrightarrow -\infty),( \Rightarrow -(-\infty)\longrightarrow +\infty)$ &  I\\
		ieee-infinity-21-22 & $(\Rightarrow +\infty^{-1}  \longrightarrow +0.0),(\Rightarrow -\infty^{-1}  \longrightarrow -0.0)$ & I \\
		ieee-infinity-24-25 & $(u=-\infty \Rightarrow u+(+\infty)\longrightarrow \texttt{NaN}), (u=-\infty \Rightarrow +\infty+u\longrightarrow \texttt{NaN})$ &  I\\
		ieee-infinity-26-27 & $(u\neq-\infty\land u\neq\texttt{NaN} \Rightarrow +\infty + u\longrightarrow+\infty),  (u\neq-\infty\land u\neq\texttt{NaN} \Rightarrow u+(+\infty)\longrightarrow+\infty)$ &  I\\
		ieee-infinity-31-32 & $(u\neq+\infty\land u\neq\texttt{NaN} \Rightarrow -\infty + u\longrightarrow-\infty), u\neq+\infty\land u\neq\texttt{NaN} \Rightarrow u+(-\infty)\longrightarrow-\infty$ &  I\\
		ieee-infinity-34-35-37-38 & $(u > 0\land u \neq \texttt{NaN} \Rightarrow u\cross +\infty\longrightarrow+\infty), (u < 0\land u \neq \texttt{NaN} \Rightarrow u\cross+\infty\longrightarrow-\infty), ( u >0 \land u\neq\texttt{NaN} \Rightarrow +\infty \cross u\longrightarrow+\infty), (u<0\land u\neq\texttt{NaN} \Rightarrow +\infty \cross u \longrightarrow-\infty)$ &  I\\
		ieee-infinity-36-39-43-46 & $(u=0 \Rightarrow u\cross+\infty\longrightarrow\texttt{NaN}), (u=0 \Rightarrow +\infty\cross u\longrightarrow\texttt{NaN}), (u=0 \Rightarrow u\cross-\infty\longrightarrow\texttt{NaN}), (u=0\Rightarrow -\infty\cross u\longrightarrow\texttt{NaN}) $&  I\\
		ieee-infinity-41-42-44-45 & $(u>0\land u\neq\texttt{NaN} \Rightarrow u\cross-\infty\longrightarrow-\infty) ,(u<0\land u\neq\texttt{NaN} \Rightarrow u\cross-\infty\longrightarrow+\infty) , (u>0\land u\neq\texttt{NaN} \Rightarrow -\infty\cross u\longrightarrow-\infty) ,(u < 0 \land u\neq\texttt{NaN} \Rightarrow -\infty\cross u\longrightarrow+\infty)$&  I\\
		ieee-infinity-47 & $(\Rightarrow +0.0^{-1} \longrightarrow +\infty), (\Rightarrow-0.0^{-1} \longrightarrow -\infty)$&  I\\
		isNaN-add-1 & $u_2 \neq (\texttt{NaN} \land \pm\infty)\Rightarrow ( u_1 + u_2 ) = \texttt{NaN}  \longrightarrow u_1 = \texttt{NaN}$ & III \\
		isNaN-add-2 & $u_1 \neq (\texttt{NaN} \land \pm\infty) \Rightarrow ( u_1 + u_2 ) = \texttt{NaN}  \longrightarrow u_2 = \texttt{NaN} $& III \\
		isNaN-div-1 & $u_2 \neq (\pm0.0\land \texttt{NaN} \land \pm\infty) \Rightarrow(u_1 / u_2) = \texttt{NaN}   \longrightarrow u_1 = \texttt{NaN}$ & III \\
		isNaN-div-2 & $u_1 \neq (\pm0.0\land \texttt{NaN} \land \pm\infty) \Rightarrow (u_1 / u_2) = \texttt{NaN} \longrightarrow u_2 = \texttt{NaN}$& III \\
		isNaN-mul-1 &$ u_2 \neq (\texttt{NaN} \land \pm\infty\land \pm0.0)  \Rightarrow ( u_1 \cross u_2 ) = \texttt{NaN} \longrightarrow u_1 = \texttt{NaN} $&  III\\
		isNaN-mul-2 & $u_1 \neq (\texttt{NaN} \land \pm\infty\land \pm0.0) \Rightarrow ( u_1 \cross u_2 ) = \texttt{NaN} \longrightarrow u_2 = \texttt{NaN}$&  III\\
		isNaN-sub & $u_2 \neq (\texttt{NaN} \land \pm\infty) \Rightarrow ( u_1 - u_2 ) = \texttt{NaN}  \longrightarrow u_1 = \texttt{NaN}$& III \\
		isNaN-fma & $u_1= \texttt{NaN}\lor u_2= \texttt{NaN}\lor u_3= \texttt{NaN} \Rightarrow \texttt{fma}(u_1, u_2, u_3) \longrightarrow \texttt{NaN}$ &  III\\
		mul-div-rem-1 & $\Rightarrow(-1.0) \cross u \longrightarrow -u$ & II \\
		mul-div-rem-1-2 & $\Rightarrow(u \cross (-1.0) \longrightarrow -u$ &  II\\
		mul-div-rem-2 &$  \texttt{RNE} \Rightarrow (-0.0 - u_1) \cross (-0.0 - u_2) \longrightarrow u_1 \cross u_2$ & II \\
		mul-div-rem-3 &$ -0.0 - (-0.0 \cross u) \longrightarrow -0.0 \cross u$& II \\
		simplify-1 & $(\Rightarrow u + (-0.0) \longrightarrow u), (\Rightarrow  u + (+0.0) \longrightarrow u) $& II \\
		simplify-1-2 & $(\Rightarrow -0.0 + u \longrightarrow u), (\Rightarrow +0.0 + u \longrightarrow u)$ & II  \\
		simplify-4 & $(u_1 = +0.0 \lor u_1 = -0.0)\land u_2\neq \pm\infty\land u_2\neq\texttt{NaN} \Rightarrow u_2 + (u_1 - u_2) \longrightarrow +0.0 $&  II\\
		simplify-7 & $(\Rightarrow u - (+0.0) \longrightarrow u),( \Rightarrow u -( -0.0) \longrightarrow u)$ &  II\\
		simplify-10 & $\Rightarrow -0.0 - (-0.0 - u) \longrightarrow u$&  II\\
		simplify-13 & $u \neq \texttt{NaN}\land u \neq \pm\infty \Rightarrow (u - u) \longrightarrow +0.0$ & II \\
		simplify-14 & $\Rightarrow u \cross 1.0 \longrightarrow u $& II \\
		simplify-14-2 & $\Rightarrow 1.0 \cross u  \longrightarrow u$ & II \\
		simplify-18 & $u\neq(\texttt{NaN}\land \pm \infty\land \pm 0.0) \Rightarrow u/u \longrightarrow 1.0 $& II \\
		simplify-19 & $(u_1=+0.0 \lor u_1= -0.0)\land u_2 \neq (\texttt{NaN}\land \pm 0.0\land \pm\infty) \Rightarrow (u_2 / u_1-u_2) \longrightarrow -1.0$ & II \\
		true-false &$ \Rightarrow (\texttt{fp.isNaN}~(\_~\texttt{NaN}~[eb]~[sb])) \longrightarrow \texttt{True}$&  III\\
		isInfinite-div &$ u_1 \neq \pm\infty \land (u_1 > 9.99999940e-01 \lor u_1 > 9.9999999999999989e-01) \Rightarrow (\texttt{fp.isInfinite}~u_2/u_1)\longrightarrow (\texttt{fp.isInfinite}~u_2)$ &  III\\
		isInfinite-add-1 & $u_1 \neq \pm\infty\land u_1 > 0\land u_2 < 0 \Rightarrow u_2 + u_1= \pm\infty \longrightarrow (\texttt{fp.isInfinite}~u_2)$ &  III\\
		isInfinite-add-2 & $u_1 \neq \pm\infty\land u_1< 0\land u_2 > 0
		\Rightarrow u_2 + u_1= \pm\infty \longrightarrow (\texttt{fp.isInfinite}~u_2) $& III \\
		isInfinite-sub-1 & $u_1\neq \pm\infty\land u_1> 0\land u_2 > 0 \Rightarrow u_2 - u_1= \pm\infty \longrightarrow (\texttt{fp.isInfinite}~u_2) $&  III\\
		isInfinite-sub-2 & $u_1\neq \pm\infty\land u_1< 0\land u_2 < 0 \Rightarrow u_2 - u_1= \pm\infty \longrightarrow (\texttt{fp.isInfinite}~u_2) $&  III\\
		isZero-mul-1 & $u_2 \neq (\texttt{NaN} \land \pm\infty)  \land u_2<0 \land  (u_1 = +0.0 \lor u_1 = -0.0) \Rightarrow u_1 \cross u_2 \longrightarrow -u_1$ &  III\\
		isZero-mul-2 & $u_1 \neq (\texttt{NaN} \land \pm\infty)  \land u_1 <0 \land  (u_2 = +0.0 \lor u_2 = -0.0) \Rightarrow u_1 \cross u_2 \longrightarrow -u_2 $&  III\\
		isZero-mul-3 & $u_2 \neq (\texttt{NaN} \land \pm\infty)  \land u_2>0 \land  (u_1 = +0.0 \lor u_1 = -0.0) \Rightarrow u_1 \cross u_2 \longrightarrow u_1$ &  III\\
		isZero-mul-4 & $u_1 \neq (\texttt{NaN} \land \pm\infty)  \land u_1 >0 \land  (u_2 = +0.0 \lor u_2 = -0.0) \Rightarrow u_1 \cross u_2 \longrightarrow u_2 $&  III\\
		isZero-div-1 &$u_2 \neq (\texttt{NaN} \land \pm\infty\land \pm 0.0) \land u_2<0 \land  (u_1 = +0.0 \lor u_1=-0.0 )\Rightarrow u_1 / u_2 \longrightarrow -u_1$& III \\
		isZero-div-2 &$u_2 \neq (\texttt{NaN} \land \pm\infty\land \pm 0.0) \land u_2>0 \land  (u_1 = +0.0 \lor u_1=-0.0 )\Rightarrow u_1 / u_2 \longrightarrow u_1$& III \\
		isNaN-sqrt-2 &$ u \neq(\texttt{NaN} \land \pm \infty)\land (u=0.0 \lor u>0) \Rightarrow \texttt{sqrt}(u) \neq \texttt{NaN}$&  III\\
		isNaN-sqrt-1 & $(u = 0 \lor u > 0) \Rightarrow \texttt{sqrt(u)} = \texttt{NaN} \longrightarrow  u =\texttt{NaN}$ &  III\\
		\bottomrule
	\end{longtable}
} 

\subsection{Relative Differences in Solving-Time for \textsc{GGen-FP}}
\label{sec:appendix-2}
\begin{figure}[t]
	\centering
	\begin{minipage}{\textwidth}
		\begin{subfigure}[b]{0.5\textwidth}
			\includegraphics[width=\linewidth]{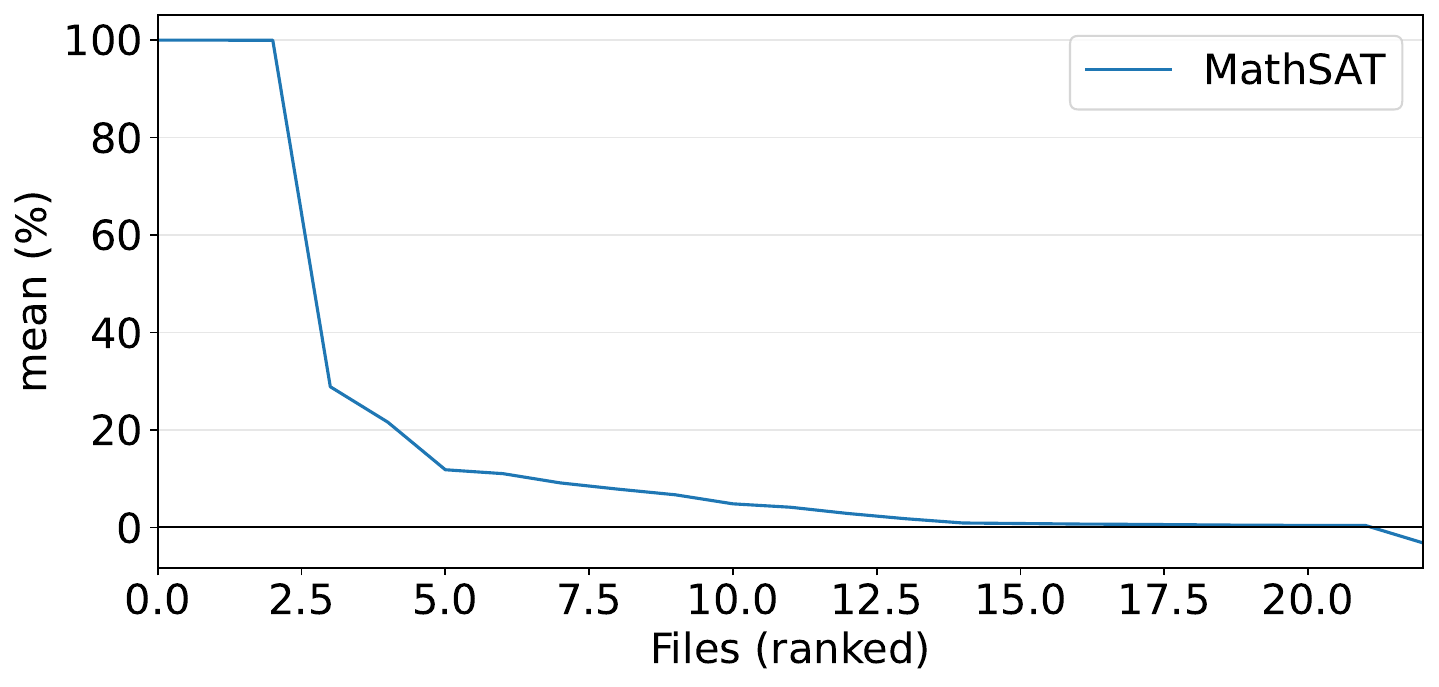}
		\end{subfigure}\hfill
		\begin{subfigure}[b]{0.5\textwidth}
			\includegraphics[width=\linewidth]{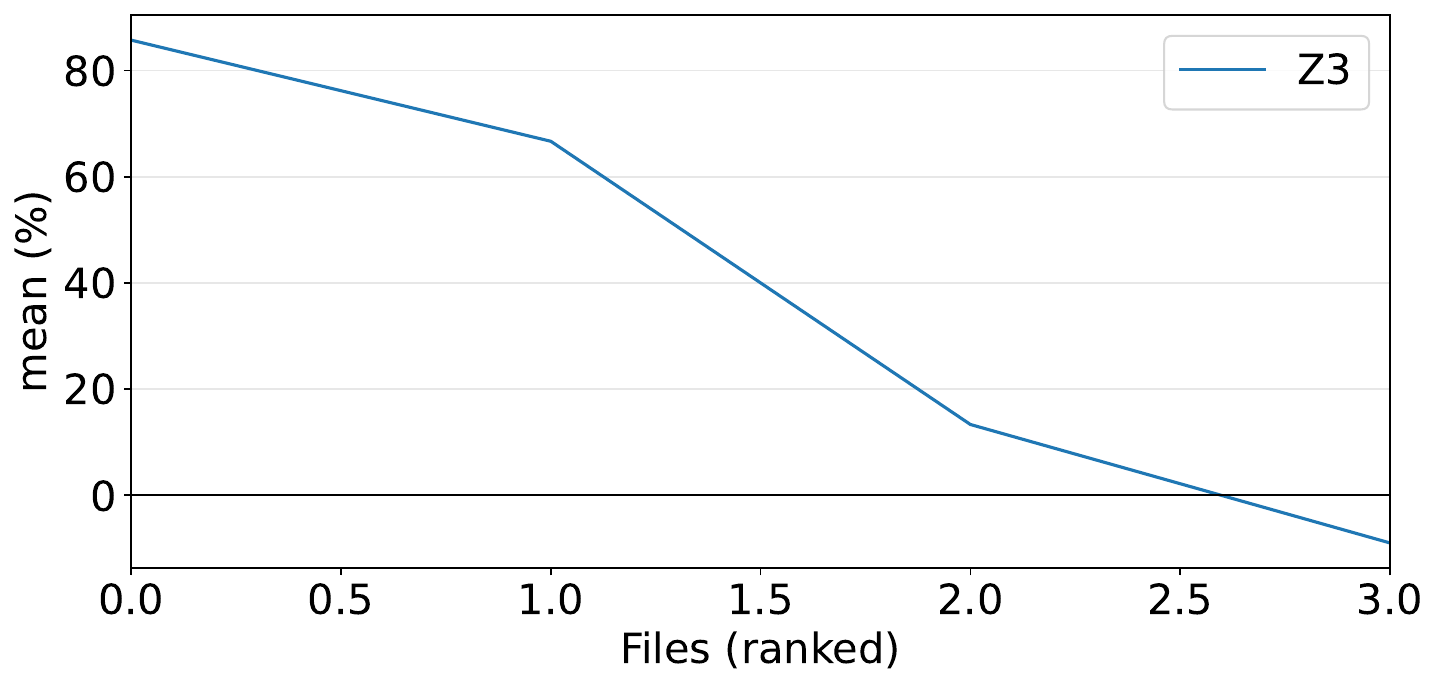}
		\end{subfigure}\hfill
		\begin{subfigure}[b]{0.5\textwidth}
			\includegraphics[width=\linewidth]{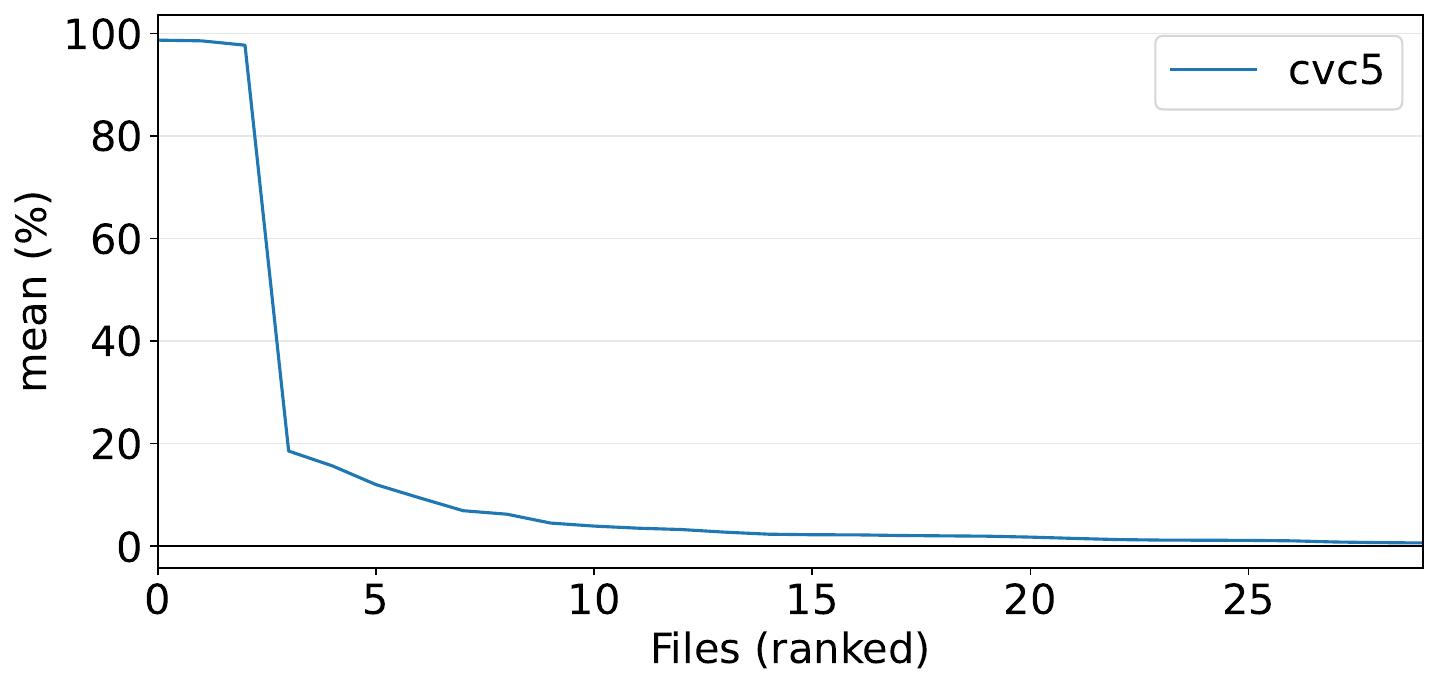}
		\end{subfigure}\hfill
		\begin{subfigure}[b]{0.5\textwidth}
			\includegraphics[width=\linewidth]{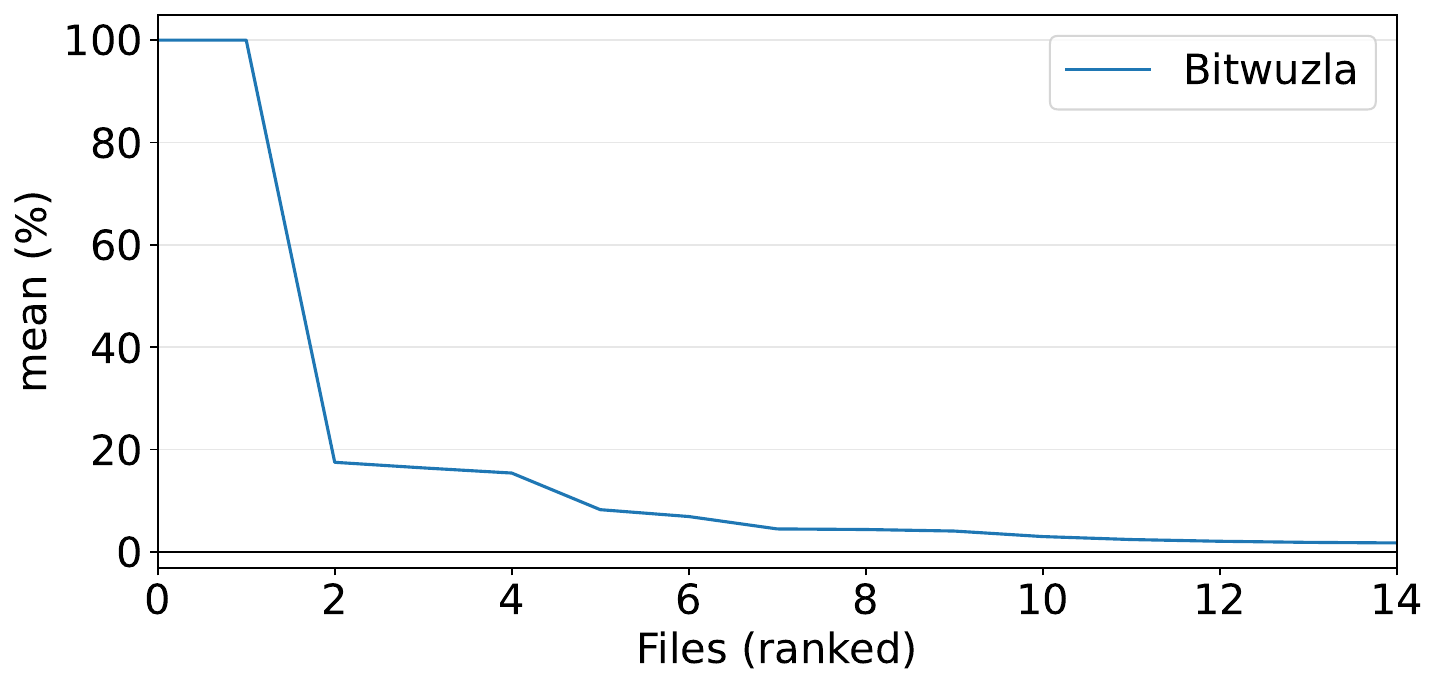}
		\end{subfigure}
		\vspace{-15pt}
		\caption{Relative difference in solving time after rewriting, \textsc{GGen-FP} suite}
		\label{fig:gentests-rel-improvment}
	\end{minipage}
\end{figure}
Figure~\ref{fig:gentests-rel-improvment} shows the magnitude of the relative
differences in solving time, for each solver sorted from the largest speedup (positive mean) to
the greatest slowdown (negative mean), for the  \textsc{GGen-FP} set. The plots only include testable files that had a
significant difference for each individual solver (so the number of files per solver differs), and show the average over the
five runs for each file.

\subsection{Effect of Rewrite Rules Independent of Normalization Assertions}
\label{sec:appendix-3}
\autoref{fig:keyfloat-rewrite-only-rel-improvment} shows the magnitude of the
relative
differences in solving time after removing the extra assertions introduced
during
normalization from the rewritten files, for each solver sorted from the largest
speedup (positive mean) to the greatest slowdown (negative mean). The plots
only include testable files that had a significant difference for each
individual
solver, and show the mean over the five runs for each file.
\begin{figure}[tbp]
	\centering
	\begin{minipage}{\textwidth}
		
		\begin{subfigure}[b]{0.5\textwidth}
			\includegraphics[width=\linewidth]{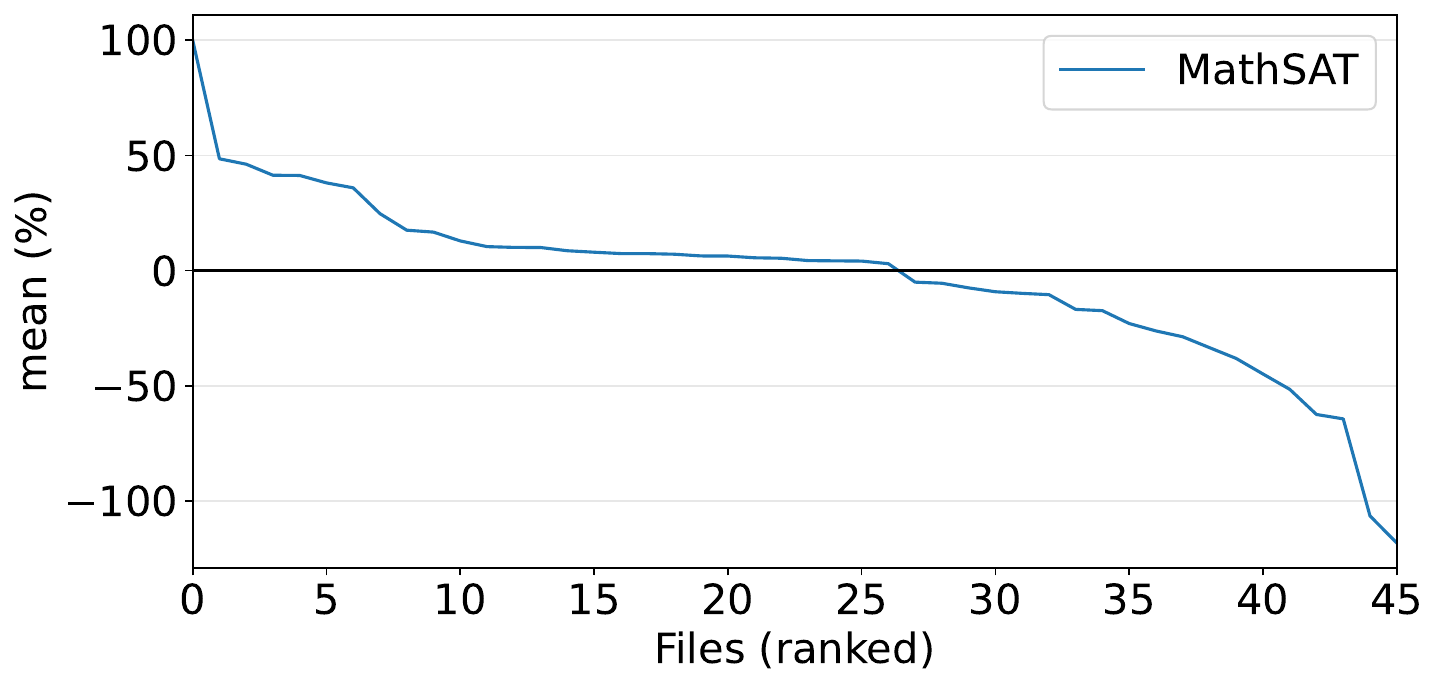}
		\end{subfigure}\hfill
		\begin{subfigure}[b]{0.5\textwidth}
			\includegraphics[width=\linewidth]{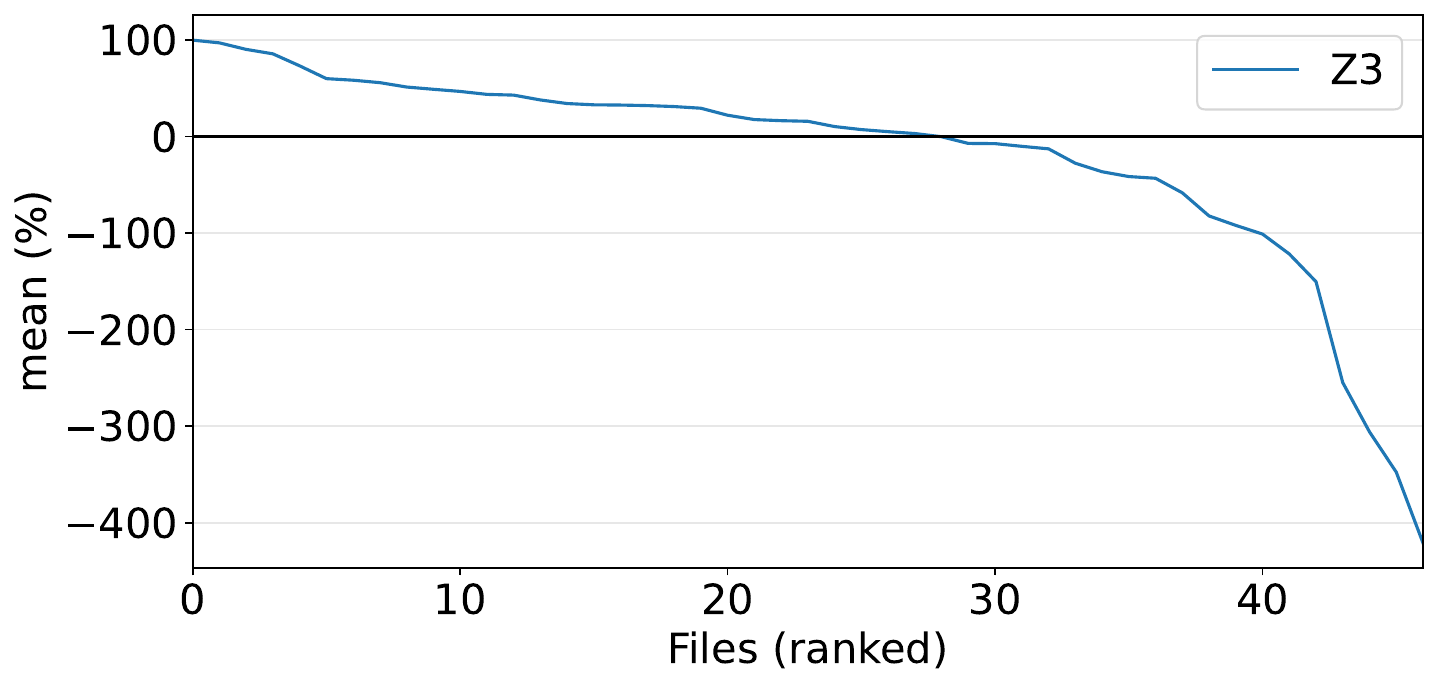}
		\end{subfigure}
		\begin{subfigure}[b]{0.5\textwidth}
			\includegraphics[width=\linewidth]{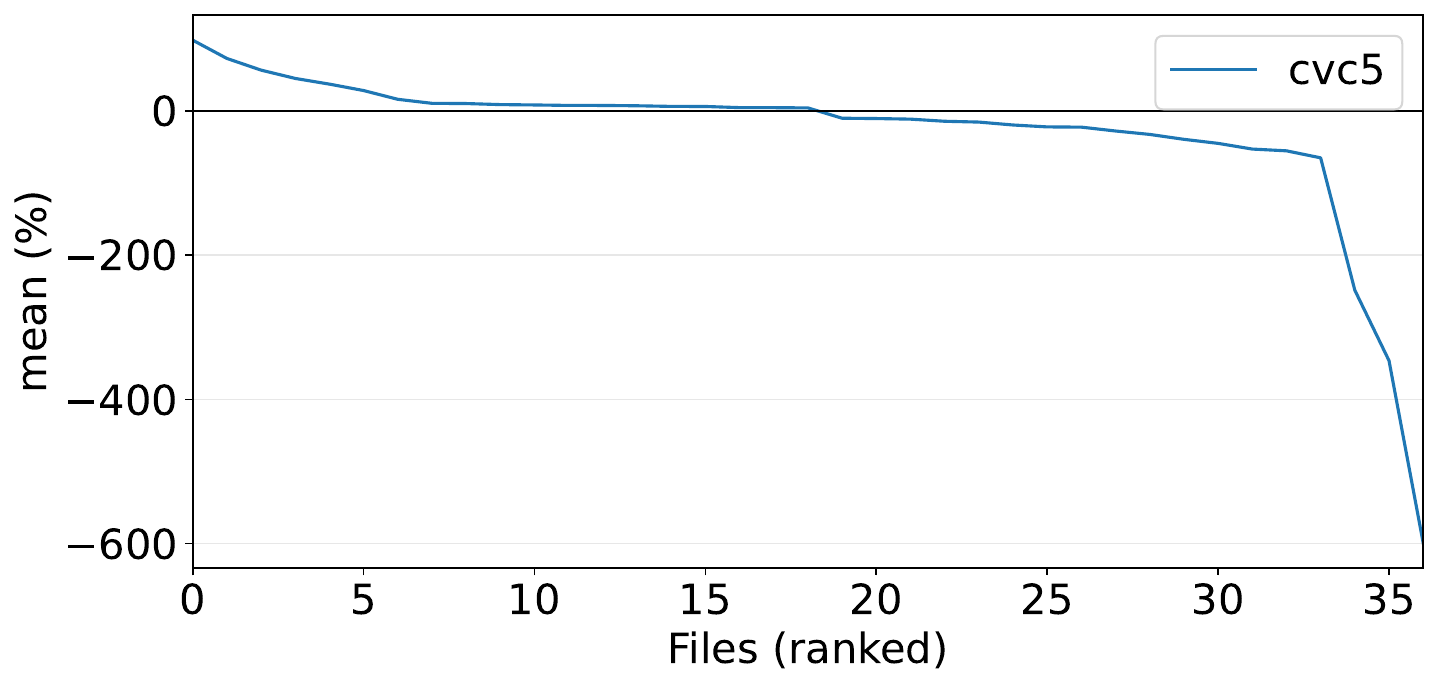}
		\end{subfigure} \hfill
		\vspace{-10pt}
		\caption{Relative difference in solving time after removing extra assertions introduced during normalization, \textsc{KeY-FP} suite}
		\label{fig:keyfloat-rewrite-only-rel-improvment}
	\end{minipage}
\end{figure}

\subsection{Effect of Normalization on Solvers' Performance}
\label{sec:appendix-4}
\begin{figure}[h] 
	\centering
	\begin{minipage}{\textwidth}
		
		\begin{subfigure}[b]{0.5\textwidth}
			\includegraphics[width=\linewidth]{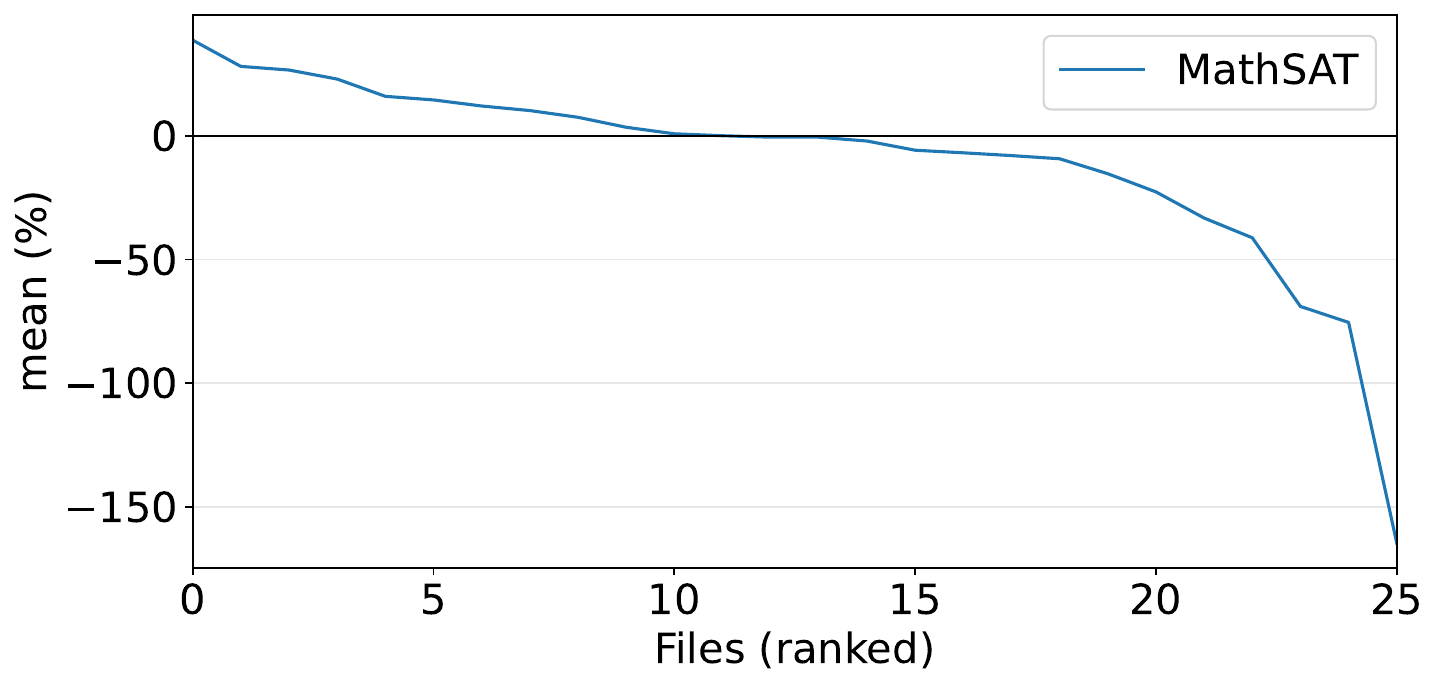}
		\end{subfigure}\hfill
		\begin{subfigure}[b]{0.5\textwidth}
			\includegraphics[width=\linewidth]{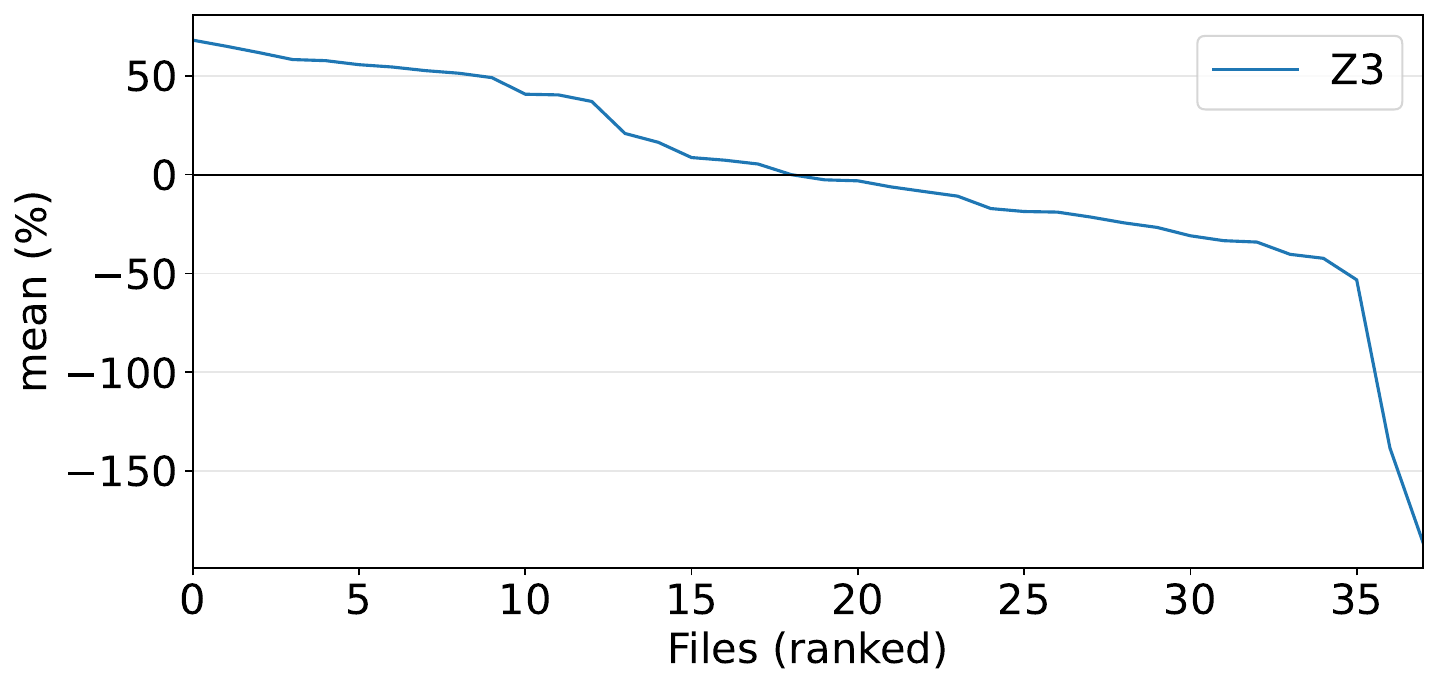}
		\end{subfigure}
		\begin{subfigure}[b]{0.5\textwidth}
			\includegraphics[width=\linewidth]{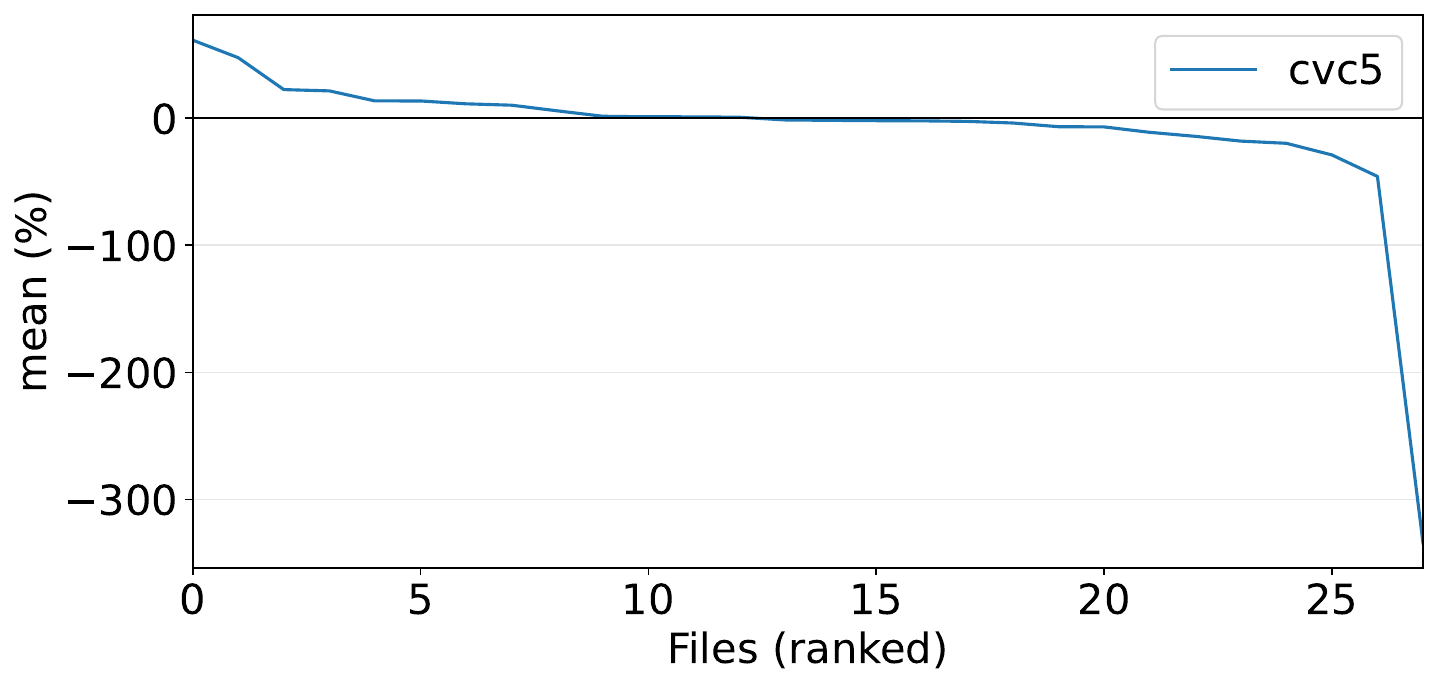}
		\end{subfigure} \hfill
		\vspace{-10pt}
		\caption{Relative difference in solving time after only applying normalization, \textsc{KeY-FP} suite}
		\label{fig:keyfloat-normal-only-rel-improvment}
	\end{minipage}
\end{figure}

\autoref{fig:keyfloat-normal-only-rel-improvment} shows the magnitude of the relative
differences in solving time, for each solver sorted from the largest speedup (positive mean) to
the greatest slowdown (negative mean). The plots only include testable files that had a
significant difference for each individual solver, and show the mean over the
five runs for each file. 
In comparison to the experiment where both rewrite rules and normalization were applied,
the number of files with significant difference is much smaller,
and the magnitude of the relative differences has also decreased
substantially.

\subsection{Impact of Full Rewriting on Total Runtime}
\label{sec:appendix-5}
Table~\ref{tab:pos-rel-ddif-summary} summarizes the positive relative differences, when applying the full set of rewrite rules to the \textsc{KeY-FP} and \textsc{KeYMut-FP} benchmark suites.

\newpage
\begin{table}[H]
	\centering
	\renewcommand{\arraystretch}{1.2}
	\caption{Summary of the positive relative differences, comparing original solving time against overall time}
	\begin{tabular}{lllll}
		\toprule
		\multirow{2}{*}{Solver} & \multicolumn{2}{c}{KeY-FP}              & \multicolumn{2}{c}{KeYMut-FP} \\
		& Mean.              & Med.               & Mean.         & Med.          \\ \hline
		Z3                     & 33.96              & 26.39             & 52.89         & 48.62        \\
		cvc5                   & 17.76              & 10.99              & 30.47        & 20.74         \\
		MathSAT~~                & 20.95              & 14.33              & 23.78         & 25.94         \\
		Bitwuzla               & \texttt{N/A} & \texttt{N/A}~~~~~  & 15.30         & 15.13        \\
		\bottomrule
		
	\end{tabular}
	\label{tab:pos-rel-ddif-summary}
\end{table}

\else

\fi
%
%

\end{document}